\documentclass[twocolumn,superscriptaddress,floatfix,prb,amsmath,amssymb,aps,showpacs]{revtex4}
\usepackage{graphicx}
\usepackage{bm}
\usepackage[dvips]{color}

\begin{document}
\bibliographystyle{apsrev}

\def\nn{\nonumber}
\def\dag{\dagger}
\def\u{\uparrow}
\def\d{\downarrow}
\def\Pf{{\rm Pf}}

\title{Projective studies of spin nematics
in a quantum frustrated ferromagnet}
\date{\today}
\author{Ryuichi Shindou}
\affiliation{Condensed Matter Theory Laboratory, RIKEN,
2-1 Hirosawa, Wako, Saitama 351-0198, Japan}
\affiliation{Physics Department, Tokyo Institute of Technology,
2-12-1 Ookayama, Meguro-ku, Tokyo 152-8551, Japan}
\author{Seiji Yunoki}
\affiliation{Computational Condensed Matter Laboratory, RIKEN ASI,
2-1 Hirosawa, Wako, Saitama 351-0198, Japan}
\affiliation{Computational Materials Science Research Team, RIKEN AICS,
Kobe, Hyogo 650-0047, Japan}
\affiliation{CREST, Japan Science and Technology Agency (JST), Kawaguchi,
Saitama 332-0012, Japan}
\author{Tsutomu Momoi}
\affiliation{Condensed Matter Theory Laboratory, RIKEN,
2-1 Hirosawa, Wako, Saitama 351-0198, Japan}
\begin{abstract}
We study the ground state properties
of the spin-$\frac12$ frustrated ferromagnetic
$J_1$--$J_2$ Heisenberg model on the square lattice,
employing projected BCS wavefunctions with
spin-triplet pairings of the spinon fields as trial
wavefunctions. Based on the variational Monte
Carlo analysis, we argue that, in the competing
coupling regime, a certain type
of the projected BCS wavefunction,
dubbed the projected $Z_2$ planar state,
achieves the best optimal energy among the other
competing states such as the ferromagnetic state and
collinear antiferromagnetic state. Like in quantum spin liquids,
the projected $Z_2$ planar state preserves the
translational symmetry of the square lattice. However,
it is also accompanied by a $d$-wave ordering of
the quadrupole moments, breaking
the spin rotational
symmetry.  
The state thus describes a quantum spin
analogue of the nematic liquid crystals.
The calculated static correlation
functions also reveal that the projected $Z_2$ planar state has
a strong collinear antiferromagnetic fluctuation.
\end{abstract}
\maketitle
\section{introduction}
Deciphering unidentified states of
quantum zero-point motion -- quantum vacuum --
is one of the central research
field in condensed matter physics.~\cite{wen}
In the realm of quantum magnetism,
investigations of {\it spin-rotational symmetric}
quantum spin liquid (QSL) -- a quantum magnet
which remains totally disordered
all the way down to the zero temperature --
belong to this research category. Indeed, after
seminal proposal of spin-singlet resonating
valence bond (RVB) wavefunctions
by Anderson and his co-workers,~\cite{RVB-paper}
tremendous research efforts
are devoted to establishing a true realization of
QSL or RVB-type ground state in spatial dimension
greater than one.~\cite{balents-Lee}
Among others, `frustrated' Mott insulating
magnets are regarded as promising
candidate materials for this investigation,~\cite{misguich}
where competing magnetic interactions between localized
spins often make it hard for a system to fall into a
simple classical spin ordering.

This work reports a variational study of spin-$\frac12$
quantum frustrated {\it ferromagnetic} model.
One motivation of
this research is a couple of experimental
works on
two-dimensional films of
solid $^3$He,~\cite{ishida,masutomi}
where the interactions between $S=\frac12$
nuclear spins of $^3$He atoms are highly
frustrated but predominantly
ferromagnetic.~\cite{roger} The
specific heat measurement~\cite{ishida} and
magnetic susceptibility~\cite{masutomi} at
the ultra-low temperature regime conclude
that the ground state of this frustrated quantum
spin system
is a QSL-like state with either
a gapless spin excitation or an extremely
small spin gap.
These experiments suggest a possibility
of an exotic quantum phase in quantum
frustrated {\it ferromagnets}.

The other incentive of this work stems from
recent theoretical studies of frustrated magnets with ferromagnetic
nearest-neighbor interactions on the square lattice 
and triangular lattice 
that report possible realization of
spin {\it nematic} phases.
\cite{Andreev1984,sms,
pre2,pre3,MomoiS,MomoiSS2006}
Among frustrated ferromagnets,
the spin-$\frac12$ square lattice $J_1$--$J_2$
model with ferromagnetic $J_1$
is a prototype minimal model
and attracting interest recently.\cite{sms,pre2,pre3,sm,pre-1,pre0,pre01,pre1,pre4,rdsfr}
The model Hamiltonian
\begin{eqnarray}
H = J_1 \sum_{\langle {\bm j},{\bm m}\rangle}
{\bm S}_{\bm j}\cdot {\bm S}_{\bm m}
+ J_2 \sum_{\langle\langle {\bm j},{\bm m}\rangle\rangle}
{\bm S}_{\bm j} \cdot {\bm S}_{\bm m} \label{hamiltonian}
\end{eqnarray}
consists of nearest-neighbor
ferromagnetic exchange $J_1$~($<0$)
and competing next-nearest-neighbor antiferromagnetic
exchange $J_2$~($>0$).
When the antiferromagnetic coupling
is much stronger than that of the ferromagnetic
coupling ($|J_1| \ll 2 J_2$),
the ground state exhibits a collinear antiferromagnetic order,
$\langle {\bm S}_{\bm j}\rangle = (-1)^{j_x} {\bm m}$ or
$(-1)^{j_y} {\bm m}$ with
${\bm j}= (j_x,j_y)$.~\cite{OrderByDisorder-paper}
While the ground state in
the opposite limit ($|J_1| \gg 2 J_2$) is the
fully polarized ferromagnetic state.
The preceding exact diagonalization studies\cite{sms,pre2,pre3} suggest
the existence of the spin nematic phase
between these two magnetic ordered phases.


Motivated by these studies, the present authors
recently formulated a fermionic mean-field
theory for quantum frustrated ferromagnets,\cite{sm}
where they proposed to describe the quantum spin
nematic phase as a {\it spin-triplet variant} of
spin-rotational symmetric quantum spin
liquids.  The state suggested
by this mean-field theory is a `mixed'
resonating valence bond (RVB) wavefunction,
where all the half-spins in
any lattice points belong to either the singlet
valence bonds on antiferromagnetic links
or the spin-triplet valence bonds introduced
on ferromagnetic links.
The wavefunction is given by a
superposition of different partitioning of spins into
either singlet or triplet bonds, such that the wavefunction,
on the whole, has no preference for any specific
valence bond configuration.

Unlike N\'eel-ordered states or dimer states, this mixed RVB state
preserves the lattice-translational symmetry of the square lattice,
indicating its `quantum spin liquid' like character.
In contrast to the spin-singlet RVB state,
however, the spin-1 moment in the triplet valence
bond breaks spin-rotational symmetry,
quantum-mechanically
rotating within a specific plane.
Thus, unlike
{\it spin-rotational symmetric}
quantum spin liquids, this
mixed RVB state is accompanied by the
breakdown of global spin-rotational symmetry.
In fact, this spontaneous symmetry breaking
manifests itself as the
ordering of the quadrupole
moment,~\cite{
Andreev1984,chubukov,sms,sm}
\begin{eqnarray}
K^{\mu\nu}_{{\bm j},{\bm m}}
\equiv \frac12 (S_{{\bm j},\mu} S_{{\bm m},\nu}
+S_{{\bm j},\nu} S_{{\bm m},\mu} )
- \frac{\delta_{\mu\nu}}{3} 
\langle {\bm S}_{{\bm j}} \cdot
{\bm S}_{{\bm m}}\rangle,
\label{n-nematic}
\end{eqnarray}
instead of the ordering of
the spin dipole moment, i.e., $\langle {\bm S}_{{\bm j}} \rangle = \textbf{0}$.
In Eq.~(\ref{n-nematic}), $\bm j$ and $\bm m$
denote two adjacent lattice sites
connected by a ferromagnetic link, and $\mu$ and $\nu$
are the spin indices. Having both
quantum spin liquid character and also symmetry breaking phase
character, this mixed RVB state is
regarded as the {\it quantum spin}
analogue of `liquid-crystal' like
state of matter.~\cite{ccl,Andreev1984,chubukov,sms,sm}

To investigate the nature of this unconventional
quantum spin state,
we study, in this paper,
the spin-$\frac12$ square lattice
$J_1$--$J_2$ Heisenberg model, using the variational Monte Carlo method.
Our method is based on the fermionic representation of
spin-$\frac12$ operators $S_{{\bm j},\mu}
= \frac12 f^{\dag}_{{\bm j},\alpha}
[\sigma_{\mu}]_{\alpha\beta}
f_{{\bm j},\beta}$ ($\mu=1,2,3$),~\cite{wen,fradkin,auerbach}
where $f^{\dag}_{{\bm j},\alpha}$ denotes the fermion creation operator
at the site ${\bm j}$ with spin $\alpha$ and $\sigma_\mu$ the Pauli matrices.
The representation becomes exact, when one and
only one fermion is located at every site, i.e.,
$f^{\dag}_{{\bm j},\alpha} f_{{\bm j},\alpha}=1$.
At the mean-field level, this local constraint on fermions'
number is replaced by their coupling with the chemical
potential, so that the only the global constraint is
taken into account.

In this fermionic representation,
the exchange interaction between
two localized spins is written as a four-point
interaction, which the mean-field theory decomposes
into the pairing fields between two adjacent fermions.
The antiferromagnetic exchange interaction is
decoupled in terms of the spin-singlet
pairing fields as~\cite{singlet,wen,fradkin,auerbach}
\begin{eqnarray}
{\bm S}_{\bm i}\cdot {\bm S}_{\bm j}
&\simeq & \frac{1}{4} \Big(|\chi_{{\bm i \bm j}}|^2
+|\eta_{{\bm i \bm j}}|^2 \Big)  \!\ + \!\ {\rm const} \nn \\
&&\hspace{-1.5cm} + \!\ \frac{1}{4} \Big(
- \chi_{{\bm j \bm i}} \!\ f^{\dag}_{{\bm i},\alpha}
f_{{\bm j},\alpha} \!\
- \!\
\eta_{{\bm j \bm i}} \!\ f^{\dag}_{{\bm i},\alpha}
[i\sigma_2]_{\alpha\beta} f^{\dag}_{{\bm j},\beta}
+ \!\ {\rm h.c.} \!\ \Big), \label{anti-ferro} 
\end{eqnarray}
where $\chi_{\bm i \bm j}$ denotes the particle-hole
(excitonic) pairing field,
\begin{equation}
\chi_{\bm i \bm j} = \langle
f^{\dag}_{{\bm i},\alpha} f_{{\bm j},\alpha}\rangle,
\end{equation}
and $\eta_{\bm i\bm j}$ is
the particle-particle (Cooper) pairing field,
\begin{equation}
\eta_{\bm i \bm j} =
\langle f_{{\bm i},\alpha}
[-i\sigma_2]_{\alpha\beta} f_{{\bm j},\beta}\rangle.
\end{equation}
On the other hand, the competing ferromagnetic
exchange interaction should be decoupled
into the spin-triplet channel in its
own right,~\cite{sm}
\begin{eqnarray}
- {\bm S}_{\bm i}\cdot {\bm S}_{\bm j}
&\simeq & \frac{1}{4}
(|{\bm E}_{{\bm i \bm j}}|^2
+ |{\bm D}_{{\bm i \bm j}}|^2) + {\rm const}. \nn \\
&&\hspace{-1.8cm} +  \frac{1}{4} \sum^3_{\mu=1}
\Big(
- E_{{\bm j \bm i},\mu} \!\
f^{\dag}_{{\bm i},\alpha}
[\sigma_{\mu}]_{\alpha\beta}
f_{{\bm j},\beta} \!\ + \!\ {\rm h.c.} \Big) \nn \\
&&\hspace{-1.8cm} + \!\ \frac{1}{4}
\sum^3_{\mu=1}
\Big(
- D_{{\bm j \bm i},\mu}
\!\ f^{\dag}_{{\bm i},\alpha}
[-i\sigma_{\mu}\sigma_{2}]_{\alpha\beta}
f^{\dag}_{{\bm j},\beta} \!\ + \!\
{\rm h.c.} \Big), \label{ferro}
\end{eqnarray}
where ${\bm D}_{{\bm j \bm m}}=(D_{{\bm j \bm m},1},D_{{\bm j \bm m},2},D_{{\bm j \bm m},3})$
denotes the so-called
$d$-vector of the spin-triplet
Cooper pairing field,~\cite{helium,helium2}
\begin{equation}
D_{{\bm i \bm j},\mu}
= \langle f_{{\bm i},\alpha}
[i\sigma_{2}\sigma_{\mu}]_{\alpha\beta}
f_{{\bm j},\beta}\rangle, \label{d-tri}
\end{equation}
and ${\bm E}_{{\bm j \bm m}}=(E_{{\bm j \bm m},1},E_{{\bm j \bm m},2},E_{{\bm j \bm m},3})$
is the excitonic counterpart
of the $d$-vector,
\begin{equation}
E_{{\bm i \bm j},\mu} =
\langle f^{\dag}_{{\bm i},\alpha}
[\sigma_{\mu}]_{\alpha\beta}
f_{{\bm j},\beta}\rangle. \label{e-tri}
\end{equation}
Physically speaking, this decoupling is because the
ferromagnetic
exchange interaction between two spin halves~\cite{sm,lzn}
usually prefers the formation of their spin-triplet valence
bond, $|S=1,S_{\mu}=0\rangle$ $(\mu=1,2,3)$,
instead of the singlet valence bond. Indeed, 
the $d$-vector associated with
the triplet pairing specifies
the direction along which the spin triplet
state has the zero magnetization, i.e.
$|S=1,({\bm e}_{d}\cdot{\bm S})=0\rangle$ with ${\bm e}_d={\bm D}_{jm}/|{\bm D}_{jm}|$.
One can easily see at the mean-field level that
the triplet pairing fields in Eqs.~(\ref{d-tri}) and (\ref{e-tri})
induce the following 
quadrupole moment,~\cite{sm}
\begin{align}
\langle K^{\mu\nu}_{{\bm i},{\bm  j}}\rangle
=& -\frac{1}{2} \big(
E_{{\bm i\bm j},\mu}
E^{*}_{{\bm i \bm j},\nu} -
\frac{1}{3}\delta_{\mu\nu}
|{\bm E}_{{\bm i\bm j}}|^2 \big)  \nn \\
& - \frac{1}{2} \big( D_{{\bm i \bm j},\mu}
D^{*}_{{\bm i \bm j},\nu}
- \frac{1}{3}\delta_{\mu\nu}
|{\bm D}_{{\bm i \bm j}}|^2\big)\ +\ {\rm h.c.}. \nn
\end{align}

The previous fermionic
mean-field analysis shows that
the present square-lattice
model, Eq.~(\ref{hamiltonian}), has
five different spin-triplet pairing
states as its saddle point solutions:~\cite{comment}
(i) $Z_2$ planar state, (ii) $Z_2$ polar state,
(iii) $SU(2)$ chiral $p$-wave
state and (iv) `flat-band' state, where the first
state is accompanied by a `coplanar' or
`d-wave' configuration of the quadrupole
moments while the second and third
ones support `collinear' configurations
of the quadrupole moments.
Among them, $Z_2$ planar phase and
$SU(2)$ chiral $p$-wave phase appear,
having the lowest energy,
in the finite range of competing coupling
regime between the ferromagnetic phase and $\pi$-flux phase
in the mean-field phase diagram.

\begin{figure}
    \includegraphics[width=8cm]{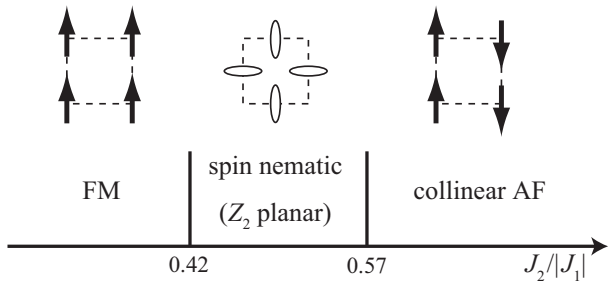}
\caption{Phase diagram of the spin-1/2 square lattice $J_1$--$J_2$ model
with ferromagnetic (FM) $J_1$ and antiferromagnetic (AF) $J_2$,
obtained from
variational
Monte Carlo simulations.
}
\label{fig:PD}
\end{figure}

In this paper, we investigate 
the nature and energetics of the projected
BCS wavefunctions constructed from these
mean-field pairing states.
As the local constraint of the fermion density
is not strictly observed in the mean-field theory,
the BCS wavefunctions
generally range
over the `extended' Hilbert space, which
allows double occupancy or vacancy
on a single site. To obtain
a proper trial many-body wavefunction for the spin
model, we first project these BCS wavefunctions
onto the physical spin Hilbert space,
imposing the single-fermion condition on every site.
By minimizing energies of these {\it projected}
BCS wavefunctions in a variational way,
we obtain their optimal energies. Comparing these
energies with those of the ferromagnetic state and the
collinear antiferromagnetic state, we argue that {\it only}
a projected $Z_2$ planar state becomes
most energetically favorable among the other
competing states in a finite range of the intermediate coupling
regime, $0.42|J_1|\alt J_2 \alt 0.57 |J_1|$.
Our results are summarized in Fig.\ \ref{fig:PD}.

Based on this observation, we further study the
character of the projected $Z_2$ planar state.
Specifically, we clarify the irreducible
representations of this many-body wavefunction
under the point group symmetries of the square lattice,
and argue that this state is actually accompanied
by a `$d$-wave' ordering of the quadrupole
moments. All irreducible representations and
the $d$-wave character are
totally consistent with the nature
of the spin nematic phase suggested by
the previous exact diagonalization
study.\cite{sms} This agreement in combination with
the energetics suggests that the projected
$Z_2$ planar state
is indeed realized in the intermediate
coupling regime of the
square lattice $J_1$--$J_2$ model. To give
a direct physical characterization to this
quantum spin nematic phase, we further
calculate the static correlation
functions in this projected BCS wavefunction 
in a large system size (100, 144, and 324 sites).
By that, we found that the wavefunction exhibits
strong antiferromagnetic fluctuation with the wave vectors
${\bm k}=(\pi,0)$ and $(0,\pi)$
associated with the proximate collinear antiferromagnetic phase,
though the finite size scaling suggests that,
the state
does not possess any staggered sublattice
magnetization in the thermodynamic limit.

The remaining sections of this paper are
organized as follows:
In Sec.~II, we briefly review spin-triplet
pairing states obtained by the previous fermionic
mean-field analysis.  We also extend these mean-field
solutions into the case under finite external Zeeman field.
By this extension, the projected `flat-band' state turns out to be
the fully polarized ferromagnetic state.
We also find that all the $d$-vectors
in the spin-triplet pairing states are lying
within a plane perpendicular to the applied field.
This feature guarantees the `spin-nematic'
character of the projected BCS wavefunctions
constructed from these pairing states.
In Sec.~III, we give a general expression
for the projected BCS wavefunctions having 
both spin-triplet and spin-singlet pairings.
Based on this expression, we have optimized
numerically the energies of (i)
projected $Z_2$ planar state, (ii) projected $Z_2$ polar state,
and (iii) projected $SU(2)$ chiral $p$-wave state.
In Sec.~IV, after briefly explaining the method of optimization
and Monte Carlo simulations, 
we compare their optimized energies with
other competing states such as the collinear antiferromagnetic state
and the fully polarized ferromagnetic state.
Sections~V and VI contain discussions about the
nature of the projected BCS wavefunctions.
In Sec.~V, we argue that all the
projected BCS wavefunctions studied in this paper
have a spin-nematic property, i.e., ordering of
the quadrupole moments {\it without} any ordering
of spins. We show in particular that the projected
$Z_2$ planar state 
is accompanied by a $d$-wave spatial configuration of
ordered quadrupole moments.
In Sec.~VI, we discuss the behavior of
the static correlation
functions of spins and
qudrupole moments calculated in
this projected $Z_2$ planar state.
Section~VII is devoted to the
summary and discussion.

\section{mean-field ansatz under the field}\label{sec:MF}
The $J_1$--$J_2$ frustrated ferromagnetic square lattice
model has four types of spin-triplet pairing states
as the saddle point solutions of the $SU(2)$ fermionic
mean-field theory: (i) $Z_2$ planar
state, (ii) $Z_2$ polar state, (iii) $SU(2)$ chiral $p$-wave state
and (iv) `flat-band' state,
all of which possess the same translational symmetry as
the square lattice. We describe in this section
how these triplet pairing states are deformed
under external Zeeman field. We
will see that all the $d$-vectors in the
states (i), (ii) and (iii) are restricted within
a plane perpendicular to the applied magnetic field.
Because of this arrangement,
the mean-field Hamiltonian
for these three states are invariant
under the spin $\pi$-rotation around the field,
when combined with the staggered gauge transformation
$f_{{\bm j},\alpha}
\rightarrow (-1)^{j_x+j_y} f_{{\bm j},\alpha}$.
This symmetry property actually
gives the spin-nematic character
to the corresponding projected
BCS wavefunctions (see Sec.~V).

\begin{figure}
    \includegraphics[width=5.5cm]{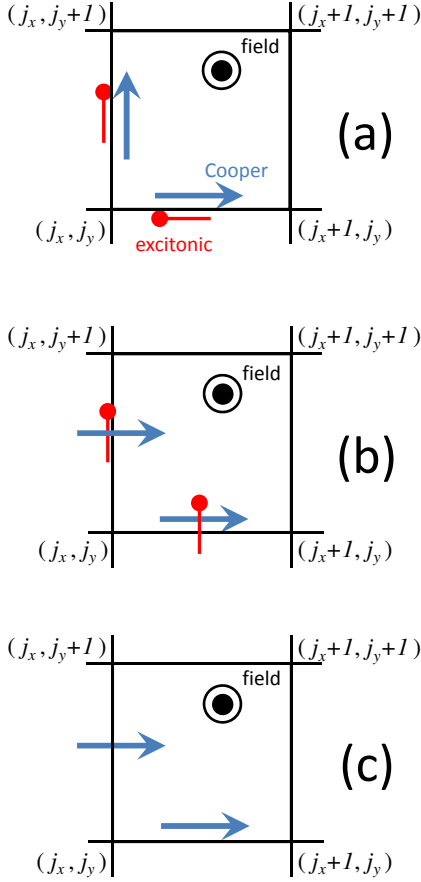}
\caption{
(Color online) A schematic picture of the
spatial configuration of $d$-vectors
in (a) $Z_2$ planar state,
(b) $Z_2$ polar state and (c) $SU(2)$
chiral $p$-wave state on the square lattice under the field.
The $d$-vectors
in the Cooper channel are drawn by (blue)
arrows and those in the excitonic
channel are given by (red) round head arrows,
both of which are on the nearest-neighbor
ferromagnetic bonds. Since all these three
states preserve the
translational symmetry of the square
lattice, we show their
configurations in the unit cell.
All the $d$-vectors are lying
within the plane perpendicular
to the applied field. These three states
are invariant under the spin $\pi$-rotation
around the field, combined with staggered gauge transformation,
which guarantees the spin-nematic character
of the corresponding projected BCS
wavefunctions (see Sec.~V).}
\label{fig:mf-uf}
\end{figure}

\subsection{$Z_2$ planar state}
In the $Z_2$ planar state in the absence of magnetic field,
the nearest-neighbor ferromagnetic bonds support
a coplanar configurations of the $d$-vector, e.g.,
\begin{subequations}
\label{Coop}
\begin{align}
D_{\bm{ij},\mu}&=\left\{\begin{array}{cc}
                  D \delta_{\mu,1}\hspace{5mm} & ({\bm i}={\bm j}+{\bm e}_x), \\
                  D \delta_{\mu,2}\hspace{5mm} & ({\bm i}={\bm j}+{\bm e}_y),
                \end{array}\right.\\
E_{\bm{ij},\mu}&=0\label{Coopb}
\end{align}
\end{subequations}
with ${\bm e}_x=(1,0)$ and ${\bm e}_y =(0,1)$, while
the next-nearest neighbor antiferromagnetic bonds
support the `staggered-flux' configurations of
the spin-singlet pairings,~\cite{ma}
\begin{equation}
\eta_{\bm{ij}}= \pm \eta,  \hspace{1cm}
\chi_{\bm{ij}}= \chi, \label{sf}
\end{equation}
when
${{\bm i}={\bm j} + {\bm e}_x \pm {\bm e}_y}$.
The corresponding Bogoliubov de-Gennes
(BdG) mean-field Hamiltonian
has the form
\begin{align}
&{\cal H}_{\rm planar}
= \sum_{{\bm j}} \bigg\{
\frac{|J_1|}{4} D \left(-f^{\dag}_{{\bm j},\alpha}
[\sigma_3]_{\alpha\beta}
f^{\dag}_{{\bm j}+{\bm e}_x,\beta}
+  i  f^{\dag}_{{\bm j},\alpha}
f^{\dag}_{{\bm j}+{\bm e}_y,\alpha}\right) \nn\\
&- \frac{J_2}{4} \sum_{\sigma=\pm} \left( \chi
f^{\dag}_{{\bm j},\alpha}
f_{{\bm j}+{\bm e}_x + \sigma {\bm e}_y,\alpha}  + i \sigma \eta
f^{\dag}_{{\bm j},\alpha}
[\sigma_2]_{\alpha\beta} f^{\dag}_{{\bm j}+{\bm e}_x+
\sigma {\bm e}_y,\beta} \right) \nn\\
&+ \!\ {\rm h.c.} \!\ \bigg\}. \label{j0}
\end{align}
This $Z_2$ planar state is energetically degenerate under spin $SU(2)$ rotation.

In the presence of magnetic field, spin $z$-component couples to both magnetic
field and mean fields from the surrounding spins. This effect
can be captured by adding the term
$-h_{\rm eff} \sum_j S_{j,z}$ to the mean-field Hamiltonian.
A direct energy optimization suggests that,
under the external Zeeman field, the coplanar plane
of the $d$-vectors is restricted
within a plane perpendicular to the field.
Moreover, the magnetic field gives rise to the excitonic triplet
pairing, whose
director vectors are also perpendicular to the field,
replacing Eq.~(\ref{Coopb}) with
\begin{align}
E_{\bm{ij},\mu}&=\left\{\begin{array}{cc}
                  -iE \delta_{\mu,1}\hspace{5mm} & ({\bm i}={\bm j}+{\bm e}_x), \\
                  iE \delta_{\mu,2}\hspace{5mm} & ({\bm i}={\bm j}+{\bm e}_y).
                \end{array}\right.
\label{exci}
\end{align}
The mean-field Hamiltonian has the
following form under the field
\begin{align}
&{\cal H}^{\prime}_{\rm planar}
= {\cal H}_{\rm planar} +
\sum_{{\bm j}} \bigg\{ \bigg[  i\frac{|J_1|}{4} E \left( f^{\dag}_{{\bm j},\alpha}
[\sigma_1]_{\alpha\beta} f_{{\bm j}+{\bm e}_x,\beta} \right.
\nn\\
& \left.
- f^{\dag}_{{\bm j},\alpha}
[\sigma_2]_{\alpha\beta} f_{{\bm j}+{\bm e}_y,\beta} \right)
+ {\rm h.c.} \bigg] -\frac12 h_{\rm eff} f^{\dag}_{{\bm j},\alpha}
[\sigma_{3}]_{\alpha\beta} f_{{\bm j},\beta} \bigg\} .\nn
\end{align}
The state preserves the translational symmetry
of the square lattice, so that the
mean-field Hamiltonian is Fourier-transformed as
\begin{align}
{\cal H}^{\prime}_{\rm planar} = 
\sum_{{\bm k}; k_y>0} {\bm f}^{\dag}_{\bm k} &\Big\{ -
\frac{|J_1|}{2} D (s_x {\bm \gamma}_3
- s_y {\bm \gamma}_5) \nn \\
&
+ \frac{|J_1|}{2} E ( s_x {\bm \gamma}_{23} 
+ s_y {\bm \gamma}_{25} )
-
J_2 \chi c_x c_y {\bm \gamma}_{4}
\nn \\
& - J_2 \eta s_x s_y {\bm \gamma}_{2}
+ \frac12 h_{\rm eff} {\bm \gamma}_{35} \Big\}
{\bm f}_{\bm k}, \label{j1}
\end{align}
where
${\bm k}=(k_x,k_y)$, ${\bm f}^{\dag}_{\bm k} \equiv
(f^{\dag}_{{\bm k},\uparrow},
f^{\dag}_{{\bm k},\downarrow},
f_{-{\bm k},\uparrow},
f_{-{\bm k},\downarrow})$, $f_{{\bm k},\alpha}\equiv \frac{1}{\sqrt{N}}
\sum_{{\bm j}} e^{i{\bm k}{\bm j}} f_{{\bm j},\alpha}$,
$s_{a} \equiv \sin k_{a}$, and
$c_{a} \equiv \cos k_{a}$ with $a=x,y$.
The $4\times 4$ ${\bm \gamma}$-matrices are
defined as ${\bm \gamma}_{1}
= {\bm \sigma}_2 \otimes {\bm \sigma}_1$,
${\bm \gamma}_{2}
= {\bm \sigma}_2 \otimes {\bm \sigma}_2$,
${\bm \gamma}_{3}
= {\bm \sigma}_2 \otimes {\bm \sigma}_3$,
${\bm \gamma}_{4}
= {\bm \sigma}_3 \otimes {\bm \sigma}_0$,
${\bm \gamma}_{5}
= {\bm \sigma}_1 \otimes {\bm \sigma}_0$,
and ${\bm \gamma}_{jm} = - i
{\bm \gamma}_j {\bm \gamma}_m$, where the
$2\times 2$ Pauli matrices in front of the
$\otimes$-mark is for the particle-hole space, while the
others are for the spin space, e.g.
\begin{eqnarray}
{\bm \gamma}_{1} =
\left(\begin{array}{cc}
0  & -i {\bm \sigma}_1  \\
i{\bm \sigma}_1 & 0 \\
\end{array}\right). \nn
\end{eqnarray}

\subsection{$Z_2$ polar state}
The $Z_2$ polar state at zero field takes a collinear configuration
of the $d$-vectors on ferromagnetic bonds, e.g.,
\begin{subequations}
\label{j-2-a}
\begin{align}
D_{\bm{ij},\mu}& = D \delta_{\mu,1},\label{j-2-aa}\\
E_{\bm{ij},\mu}&=0 \label{j-2-ab}
\end{align}
\end{subequations}
for ${\bm i}={\bm j}+{\bm e}_a$ $(a=x,y)$,
while it supports the same staggered-flux
configuration of the singlet pairings as
in Eq.~(\ref{sf}).

In the presence of magnetic field, the external Zeeman field
restricts the $d$-vectors within
its transverse directions the same as in Eq.~(\ref{j-2-aa})
and brings
about the excitonic spin-triplet pairings,
replacing Eq.~(\ref{j-2-ab}) with
\begin{equation}
E_{\bm{ij},\mu}=iE \delta_{\mu,2}
\end{equation}
for ${\bm i}={\bm j}+{\bm e}_a$ $(a=x,y)$.
The $d$-vectors of the induced excitonic spin-triplet pairings
are perpendicular to both the
external field and the $d$-vectors of the
Cooper channel.
The mean-field Hamiltonian for the
$Z_2$ polar state has
the following form under the
field,
\begin{align}
&{\cal H}_{\rm polar}
= \sum_{{\bm j}} \bigg\{ \bigg[
\alpha_1
f^{\dag}_{{\bm j},\u} f^{\dag}_{{\bm j},\d} \nn \\
&
- \frac{|J_1|}{4} \sum_{a=x,y} \Big(D f^{\dag}_{{\bm j},\alpha}
[\sigma_3]_{\alpha\beta}
f^{\dag}_{{\bm j}+{\bm e}_a,\beta}
 + i E  f^{\dag}_{{\bm j},\alpha}
[\sigma_2]_{\alpha\beta}
f_{{\bm j}+{\bm e}_a,\beta} \Big)
\nn\\
&- \frac{J_2}{4} \sum_{\sigma=\pm} \Big( \!\  \chi \!\
f^{\dag}_{{\bm j},\alpha}
f_{{\bm j}+{\bm e}_x + \sigma {\bm e}_y,\alpha}
+ i \sigma \eta
f^{\dag}_{{\bm j},\alpha}
[\sigma_2]_{\alpha\beta} f^{\dag}_{{\bm j}+{\bm e}_x+
\sigma {\bm e}_y,\beta} \!\ \Big)\nn \\
&
+ {\rm h.c.} \bigg] -\frac12 h_{\rm eff} f^{\dag}_{{\bm j},\alpha}
[\sigma_{3}]_{\alpha\beta} f_{{\bm j},\beta} \bigg\},
\end{align}
where $\alpha_1$ denotes the uniform
temporal gauge field
that has a finite value in the $Z_2$ polar state
(see Ref.~\onlinecite{sm}).
Or equivalently,
\begin{align}
{\cal H}_{\rm polar} &=
\sum_{{\bm k};k_y>0} {\bm f}^{\dag}_{\bm k} \bigg\{
-\alpha_1 {\bm \gamma}_{2}
- \frac{|J_1|}{2} (s_x + s_y )
( D {\bm \gamma}_3
- E {\bm \gamma}_{25})  \nn \\
& -  J_2 \chi c_x c_y
{\bm \gamma}_{4} -  J_2 \eta s_x s_y {\bm \gamma}_{2}
+ \frac12 h_{\rm eff} {\bm \gamma}_{35} \bigg\} {\bm f}_{\bm k}. \label{j2}
\end{align}

\subsection{$SU(2)$ chiral $p$-wave state}
In the $SU(2)$ chiral $p$-wave state, the $d$-vectors
on the nearest neighbor $x$-link and that on the $y$-link
are collinear with each other. One of the two
acquires an additional phase factor $i$,
compared with the other,
\begin{subequations}
\begin{align}
D_{\bm{ij},\mu}&=\left\{\begin{array}{cc}
                  D \delta_{\mu,1}\hspace{5mm} & ({\bm i}={\bm j}+{\bm e}_x), \\
                  iD \delta_{\mu,1}\hspace{5mm} & ({\bm i}={\bm j}+{\bm e}_y),
                \end{array}\right.\\
E_{\bm{ij},\mu}&=0.
\end{align} \label{chiral-p-1}
\end{subequations}
Eqs.~(\ref{chiral-p-1}a-b) hold under
any magnetic field,
provided that the ${\bm d}$-vector is
perpendicular to the field.
The antiferromagnetic exchange
interaction supports  the `uniform-RVB'
configuration of the singlet pairings,~\cite{singlet}
\begin{equation}
\eta_{\bm{ij}}= 0,  \hspace{1cm}
\chi_{\bm{ij}}= \chi
\end{equation}
for
${{\bm i}={\bm j} + {\bm e}_x \pm {\bm e}_y}$.
The mean-field Hamiltonian for
the $SU(2)$ chiral $p$-wave state takes the
following form,
%
\begin{align}
&{\cal H}_{{\rm chiral} \!\ p \!\ {\rm wave}}
= \sum_{{\bm j}} \bigg\{ \bigg[
 - \frac{|J_1|}{4} D \Big(f^{\dag}_{{\bm j},\alpha}
[\sigma_3]_{\alpha\beta}
f^{\dag}_{{\bm j}+{\bm e}_x,\beta} \nn \\
&
+ i  f^{\dag}_{{\bm j},\alpha}
[\sigma_3]_{\alpha\beta}
f^{\dag}_{{\bm j}+{\bm e}_y,\beta} \Big) \!\
- \frac{J_2}{4} \chi \sum_{\sigma=\pm}
f^{\dag}_{{\bm j},\alpha}
f_{{\bm j}+{\bm e}_x + \sigma {\bm e}_y,\alpha} \nn \\
&
+ {\rm h.c.} \bigg] - \frac12 h_{\rm eff} \!\  f^{\dag}_{{\bm j},\alpha}
[\sigma_{3}]_{\alpha\beta} f_{{\bm j},\beta}
\bigg\},
\end{align}
or equivalently,
\begin{align}
{\cal H}_{{\rm chiral} \!\ p \!\ {\rm wave}}
= &
\sum_{{\bm k};k_y>0} {\bm f}^{\dag}_{\bm k} \Big\{
-  \frac{|J_1|}{2} D (s_x {\bm \gamma}_3
+  s_y {\bm \gamma}_{34}) \nn \\
&  -  J_2 \chi c_x c_y
{\bm \gamma}_{4} + \frac12 h_{\rm eff} {\bm \gamma}_{35} \Big\}
{\bm f}_{\bm k}. \label{j3}
\end{align}

\subsection{Fully polarized state out of `flat-band' states}
The `flat-band' states have only spin-triplet pairings
on the ferromagnetic bonds, while no spin-singlet
pairing on the antiferromagnetic bonds. According
to our previous work,~\cite{sm} this state achieves the
best mean-field energy among others in the strongly
ferromagnetic regime ($|J_1|\gg J_2$).
In the absence of the external field, the triplet
pairings in the `flat-band' state are most generally
characterized by a $U(1)$ phase, $\theta$, three
orthogonal unit vectors $\{ {\bm n}_1, {\bm n}_2, {\bm n}_3\}$
in the spin space,
and two orthogonal unit vectors $\{{\bm m}_1, {\bm m}_2\}$
in the gauge space as follows
\begin{subequations}
\begin{align}
{\bm D}_{\bm{ij}}&=\left\{\begin{array}{cc}
\cos \theta \!\ {\bm n}_{1} (m_{1,3}
+ i  \!\ m_{1,2})\hspace{5mm} & ({\bm i}={\bm j}+{\bm e}_x), \\
\sin \theta \!\ {\bm n}_{1} (m_{2,3}
+ i \!\ m_{2,2})\hspace{5mm} & ({\bm i}={\bm j}+{\bm e}_y),
                \end{array}\right.\\
{\bm E}_{\bm{ij}}&=\left\{\begin{array}{cc}
 \cos \theta \!\ ({\bm n}_{2}
+ i \!\ {\bm n}_{1} m_{1,1})\hspace{5mm} & ({\bm i}={\bm j}+{\bm e}_x), \\
\sin \theta \!\ ({\bm n}_{3}
+ i \!\ {\bm n}_{1} m_{2,1})\hspace{5mm} & ({\bm i}={\bm j}+{\bm e}_y),
                \end{array}\right.
\end{align}
\end{subequations}
where
\begin{eqnarray}
&&{\bm n}_1 \cdot {\bm n}_2 = {\bm n}_2 \cdot {\bm n}_3
= {\bm n}_3 \cdot {\bm n}_1=0, \label{spin} \\
&&
{\bm m}_1 \cdot {\bm m}_2=0.  \label{gauge}
\end{eqnarray}
The energy dispersion of the Bogoliubov particle
comprises two bands, which are totally flat in the momentum space and
energetically separated by $2|J_1|$ from each other.
%
In the remaining part of this section,
we will argue that this state actually
reduces to a fully polarized ferromagnetic
state, once an infinitesimally small Zeeman
field is applied.

Under the Zeeman field, it is energetically
favorable that all the $d$-vectors are
perpendicular to the field.
To make this compatible with
the orthogonality condition Eq.~(\ref{spin}),
the $U(1)$ phase $\theta$ is going to be locked in
$\theta=\frac{\pi}{2}l$ with $l\in \mathbb{Z}$.
Namely, when $\theta=\pm \frac{\pi}{2}$, only
${\bm n}_3$ and ${\bm n}_1$ are
required to be perpendicular to the field,
while, in the case of $\theta=0$ or $\pi$,
only ${\bm n}_1$ and ${\bm n}_2$ are
perpendicular to the field. Thereby,
this locking reduces the `flat-band'
state into a decoupled one-dimensional
fermion states running along either
$x$-link or $y$-link. For example, when
$\theta=0$, one of the flat-band states
under an infinitesimally small field
can be described with
${\bm n}_1=(1,0,0)$,
${\bm n}_2=(0,1,0)$, and
${\bm m}_1=(1,0,0)$.
The corresponding BdG Hamiltonian
is given as
\begin{align}
{\cal H}_{\rm flat} &= \sum_{\bm j} \bigg\{
\bigg[ - \frac{|J_1|}{4}
\Big(i f^{\dag}_{{\bm j},\alpha}
[\sigma_1]_{\alpha\beta}
f_{{\bm j}+e_x,\beta} \nn \\
&+ f^{\dag}_{{\bm j},\alpha}
[\sigma_2]_{\alpha\beta}
f_{{\bm j}+e_x,\beta} \Big) + {\rm h.c.} \bigg]
- \frac12 h_{\rm eff}  f^{\dag}_{{\bm j},\alpha}
[\sigma_{3}]_{\alpha\beta} f_{{\bm j},\beta}
 \bigg\}.
\end{align}
Or equivalently,
\begin{align}
{\cal H}_{\rm flat} =  \sum_{j_y=1}^{L_y}
\sum_{-\pi<k_x<\pi} & {\bm f}^{\dag}_{k_x,j_y} \Big\{
 - \!\
\frac{|J_1|}{2} (c_x {\bm \gamma}_{31}  \!\
+ s_x {\bm \gamma}_{23} ) \nn \\
&  + \!\ \frac12 h_{\rm eff} {\bm \gamma}_{35}
 \Big\} {\bm f}_{k_x,j_y},
\label{flat-band}
\end{align}
where ${\bm f}^{\dag}_{k_x,j_y}\equiv
(f^{\dag}_{k_x,j_y,\u},f^{\dag}_{k_x,j_y,\d},
f_{-k_x,j_y,\u}, f_{-k_x,j_y,\d})$ and $
f_{k_x,j_y,\alpha} \equiv
\frac{1}{\sqrt{L}_x} \sum^{L_x}_{j_x=1}
e^{ik_x j_x} f_{{\bm j},\alpha}$
with $L_x L_y = N$.

When projected into the spin Hilbert space,
the ground-state wavefunction of Eq.~(\ref{flat-band})
reduces to a fully polarized ferromagnetic state.
To see this, 
note that it 
is given by a composite of decoupled one-dimensional
fermionic states running along the $x$-link;
$|\Psi_{\rm flat}\rangle =
\prod^{L_y}_{j_y=1}|\Psi_{j_y}\rangle$.
For every $j_y=1,\cdots,L_y$,
$|\Psi_{j_y}\rangle$ is given by
\begin{eqnarray}
|\Psi_{j_y}\rangle &=& \prod_{-\pi<k_x<\pi}
\Big( -\!\ \cos \frac{\phi}{2} \!\
e^{i\frac{\pi}{4} + i\frac{k_x}{2}}
f^{\dag}_{k_x\uparrow} \nn \\
&& \!\ + \!\ \sin \frac{\phi}{2} \!\
e^{-i\frac{\pi}{4} - i\frac{k_x}{2}}
f^{\dag}_{k_x,\downarrow} \Big) \!\ \big|0\big\rangle, \label{wferr}
\end{eqnarray}
where $|0\rangle$ denotes the vacuum state of
the fermions. We have omitted the index `$j_y$',
$f^{\dag}_{k_x,j_y,\alpha} \rightarrow
f^{\dag}_{k_x,\alpha}$, since the
argument holds for each $j_y$ independently.
Note also that the quantization axis of the spin
was taken along the field. The angle $\phi$
is defined as
\begin{eqnarray}
\phi = \tan^{-1}\left[\frac{|J_1|}{h_{\rm eff}}\right]
 \nn
\end{eqnarray}
in the range $-\frac{\pi}{2} \le \phi \le \frac{\pi}{2}$.

The inner-product between $|\Psi_{j_y}\rangle$ and
an Ising spin configuration $|\{\sigma_{j_x}\}\rangle \equiv
\{\prod^{L_x}_{j_x=1} f^{\dag}_{j_x,\sigma_{j_x}}\} |0\rangle$
is given by a determinant of the $L_x \times L_x$ matrix,
\begin{widetext}
\begin{eqnarray}
\langle \{ \sigma_{j_x} \} |\Psi_{j_y}\rangle
=
{\rm det}\left[\begin{array}{cccc}
a_{\sigma_1} e^{i\frac{\pi \sigma_1}{4}}
e^{i\frac{2\pi}{L_x}(-1+\frac{\sigma_1}{2})}
&
a_{\sigma_2} e^{i\frac{\pi \sigma_2}{4}}
e^{i\frac{2\pi}{L_x}(-2+\frac{\sigma_2}{2})}
& \cdots
& a_{\sigma_{L_x}} e^{i\frac{\pi \sigma_{L_x}}{4}}
e^{i\frac{2\pi}{L_x}(-L_x+\frac{\sigma_{L_x}}{2})} \\
a_{\sigma_1} e^{i\frac{\pi \sigma_1}{4}}
e^{i\frac{4\pi}{L_x}(-1+\frac{\sigma_1}{2})}
&
a_{\sigma_2} e^{i\frac{\pi \sigma_2}{4}}
e^{i\frac{4\pi}{L_x}(-2+\frac{\sigma_2}{2})}
& \cdots
& a_{\sigma_{L_x}} e^{i\frac{\pi \sigma_{L_x}}{4}}
e^{i\frac{4\pi}{L_x}(-L_x+\frac{\sigma_{L_x}}{2})} \\
\vdots &\vdots & \ddots & \vdots \\
a_{\sigma_1} e^{i\frac{\pi \sigma_1}{4}}
e^{i 2\pi (-1+\frac{\sigma_1}{2})} &
a_{\sigma_2} e^{i\frac{\pi \sigma_2}{4}}
e^{i 2\pi (-2+\frac{\sigma_2}{2})}
&
\cdots
& a_{\sigma_{L_x}} e^{i\frac{\pi \sigma_{L_x}}{4}}
e^{i 2\pi (-L_x+\frac{\sigma_{L_x}}{2})} \\
\end{array}\right]. \nn
\end{eqnarray}
\end{widetext}
where $\sigma=\pm 1$ ($\uparrow,\downarrow$
respectively) with
$a_{\sigma}\equiv -\delta_{\sigma,1} \cos\frac{\phi}{2}
+ \delta_{\sigma,-1} \sin \frac{\phi}{2}$.
This determinant becomes non-zero, if and only if all
the spins are pointing upward or pointing downward: otherwise,
the $L_x \times L_x$ matrix always has two adjacent
column-vectors which are parallel to each other
in the $L_x$ dimensional space,
at a domain wall with $(\sigma_l,\sigma_{l+1})=(-1,1)$.
Accordingly, we have
\begin{eqnarray}
\langle \{\sigma_{j_x}\}
|\Psi_{j_y}\rangle & =& \big(-\cos{\frac{\phi}{2}}\big)^{L_x}
\prod^{L_x}_{j_x=1} \delta_{\sigma_{j_x},+1} \nn \\
&& + \big(\sin{\frac{\phi}{2}}\big)^{L_x}
\prod^{L_x}_{j_x=1}
\delta_{\sigma_{j_x},-1}.
\end{eqnarray}
Note that
$\cos\frac{\phi}{2} > \sin\frac{\phi}{2} \ge 0$
for $h_{\rm eff} > 0$. This suggests that, when normalized in
the thermodynamic limit, Eq.~(\ref{wferr})
always reduces to the ferromagnetic state,
\begin{eqnarray}
\lim_{L_x \rightarrow \infty}
\frac{1}{\sqrt{\langle \Psi_{j_y} |\Psi_{j_y}\rangle}} \!\
|\Psi_{j_y}\rangle =
|\uparrow,\uparrow,\cdots,\uparrow \rangle \label{ferr}
\end{eqnarray}
for any $j_y=1,\cdots, L_y$.

\section{trial many-body wavefunction}
As shown in the previous section, the
projected `flat-band' state under an infinitesimally
small field reduces to the fully polarized
ferromagnetic state. As such, we regard that the
projected `flat-band' state in the absence of the field
is the trivial ferromagnetic state, whose
energy is exactly estimated as
$-\frac{1}{2}(|J_1|-J_2)$ (per site).
Hence we will focus on the character and
energetics of the other three spin-triplet RVB
states, (i) $Z_2$ planar state, (ii) $Z_2$ polar
state and (iii) $SU(2)$ chiral $p$-wave state.
To this end, we will construct in this section
the {\it projected} spin-triplet BCS wavefunctions
out of their respective mean-field Hamiltonians.


\subsection{Projected spin-triplet BCS wavefunctions}
Let us first derive a BCS `many-body' wavefunction
for the BdG Hamiltonian which has both spin-triplet and spin-singlet
pairings and hopping integrals.
Suppose that we have a mean-field Hamiltonian
${\cal H} \equiv \sum_{k_y>0}
{\bm f}^{\dag}_{\bm k} {\bm H}_{\bm k}
{\bm f}_{\bm k}$
and
a $4 \times 4$ matrix 
${\bm H}_{\bm k}$
is diagonalized by a unitary
transformation ${\bm U}_{\bm k}$.
We typically use
Eqs.~(\ref{j1},\ref{j2},\ref{j3}) for ${\cal H}$.
For these Hamiltonians, the eigenvalues
always appear in the particle-hole pairwise manner,
\begin{eqnarray}
{\bm H}_{\bm k}\!\ {\bm U}_{\bm k}
=
{\bm U}_{\bm k} \!\ \left[\begin{array}{cccc}
\lambda_{{\bm k},1} & & & \\
 & \lambda_{{\bm k},2} & & \\
 & & -\lambda_{{\bm k},2} & \\
& & & - \lambda_{{\bm k},1} \\
\end{array}\right].
\end{eqnarray}
As such, without loss of generality, we can
assume $\lambda_{{\bm k},j}$ ($j=1,2$)
to be positive (semi-)definite.
Defining the Bogoliubov
particle ${\gamma}^{(\dag)}_{{\bm k},j}$
$(j=1,2)$ as
\begin{eqnarray}
&& \left(\begin{array}{cccc}
\gamma^{\dag}_{{\bm k},1} &
\gamma^{\dag}_{{\bm k},2}
& \gamma_{-{\bm k},2} &
\gamma_{-{\bm k},1} \end{array}\right)
\nn \\
&&
\equiv
\left(\begin{array}{cccc}
f^{\dag}_{{\bm k},\u} &
f^{\dag}_{{\bm k},\d} & f_{-{\bm k},\u} &
f_{-{\bm k},\d}
\end{array}\right) \!\ {\bm U}_{\bm k} , \label{a0}
\end{eqnarray}
we obtain
\begin{eqnarray}
{\cal H}
= \sum_{j=1}^2 \sum_{{\bm k}; k_y>0}
\big\{\lambda_{{\bm k},j}
\gamma^{\dag}_{{\bm k},j} \gamma_{{\bm k},j}
- \lambda_{{\bm k},j} \gamma_{-{\bm k},j}
\gamma^{\dag}_{-{\bm k},j}\big\}.
\end{eqnarray}
Since $\lambda_{{\bm k},j}\ge 0$,
the mean-field ground state wavefunction $|{\rm g.s.} \rangle$ is
a vacuum of the Bogoliubov particles, i.e.,
$\gamma_{{\bm k},j}|{\rm g.s.} \rangle =
\gamma_{-{\bm k},j} | {\rm g.s.} \rangle =0 $ for any ${\bm k}$ and $j$,
which leads to
\begin{eqnarray}
|{\rm g.s.} \rangle \propto \prod_{{\bm k};k_y>0}
\big\{ \gamma_{{\bm k},1}
\gamma_{{\bm k},2}
\gamma_{-{\bm k},2} \gamma_{-{\bm k},1}\big\}
\big| 0 \big\rangle. \label{a1}
\end{eqnarray}
Substituting Eq.~(\ref{a0}) into Eq.~(\ref{a1}),
one can easily obtain
\begin{eqnarray}
&& \hspace{-0.6cm} \prod_{{\bm k};k_y>0}
\big\{ \gamma_{{\bm k},1}
\gamma_{{\bm k},2}
\gamma_{-{\bm k},2} \gamma_{-{\bm k},1}\big\}
\big| 0 \big\rangle \nn \\
 && \ \  = c\!\ \prod_{{\bm k};k_y>0}
\big\{1 + a_{\bm k} f^{\dag}_{-{\bm k},\u}
f^{\dag}_{{\bm k},\d}
+ b_{\bm k} f^{\dag}_{-{\bm k},\d}f^{\dag}_{{\bm k},\d}
\nn \\
&& \hspace{-0.8cm}
+ a^{\prime}_{\bm k} f^{\dag}_{-{\bm k},\u}
f^{\dag}_{{\bm k},\u}
+ b^{\prime}_{\bm k} f^{\dag}_{-{\bm k},\d}
f^{\dag}_{{\bm k},\u}
+ c_{\bm k} f^{\dag}_{-{\bm k},\u}
f^{\dag}_{-{\bm k},\d}
f^{\dag}_{{\bm k},\u} f^{\dag}_{{\bm k},\d} \big\}
\big| 0 \big\rangle  \nn \\
 \label{gs1}
\end{eqnarray}
with
\begin{eqnarray}
c &=& \prod_{k_y>0}
\big\{ - [{\bm U}_{\bm k}]_{1,3}
[{\bm U}_{\bm k}]_{2,4}
+ [{\bm U}_{\bm k}]_{2,3}
[{\bm U}_{\bm k}]_{1,4} \big\}  \nn \\
&&\hspace{0.2cm} \times \!\
\big\{[{\bm U}_{\bm k}]^{*}_{1,1}
[{\bm U}_{\bm k}]^{*}_{2,2}
- [{\bm U}_{\bm k}]^{*}_{1,2}
[{\bm U}_{\bm k}]^{*}_{2,1} \big\}
 \label{t1}
\end{eqnarray}
and
\begin{eqnarray}
a_{\bm k} &=& \frac{[{\bm U}_{\bm k}]^{*}_{1,1}
[{\bm U}_{\bm k}]^{*}_{3,2}
- [{\bm U}_{\bm k}]^{*}_{1,2}
[{\bm U}_{\bm k}]^{*}_{3,1}}
{[{\bm U}_{\bm k}]^{*}_{1,1} [{\bm U}_{\bm k}]^{*}_{2,2}
- [{\bm U}_{\bm k}]^{*}_{1,2}
[{\bm U}_{\bm k}]^{*}_{2,1}}, \nn \\
b_{\bm k} &=&
\frac{[{\bm U}_{\bm k}]^{*}_{1,1}
[{\bm U}_{\bm k}]^{*}_{4,2}
- [{\bm U}_{\bm k}]^{*}_{1,2}
[{\bm U}_{\bm k}]^{*}_{4,1}}
{[{\bm U}_{\bm k}]^{*}_{1,1}
[{\bm U}_{\bm k}]^{*}_{2,2}
- [{\bm U}_{\bm k}]^{*}_{1,2}
[{\bm U}_{\bm k}]^{*}_{2,1}}, \nn \\
a^{\prime}_{\bm k}
&=& - \frac{[{\bm U}_{\bm k}]^{*}_{2,1}
[{\bm U}_{\bm k}]^{*}_{3,2}
- [{\bm U}_{\bm k}]^{*}_{2,2}
[{\bm U}_{\bm k}]^{*}_{3,1}}
{[{\bm U}_{\bm k}]^{*}_{1,1}
[{\bm U}_{\bm k}]^{*}_{2,2}
- [{\bm U}_{\bm k}]^{*}_{1,2}
[{\bm U}_{\bm k}]^{*}_{2,1}}, \nn \\
b^{\prime}_{\bm k} &=&
- \frac{[{\bm U}_{\bm k}]^{*}_{2,1}
[{\bm U}_{\bm k}]^{*}_{4,2}
- [{\bm U}_{\bm k}]^{*}_{2,2}
[{\bm U}_{\bm k}]^{*}_{4,1}}
{[{\bm U}_{\bm k}]^{*}_{1,1}
[{\bm U}_{\bm k}]^{*}_{2,2}
- [{\bm U}_{\bm k}]^{*}_{1,2}
[{\bm U}_{\bm k}]^{*}_{2,1}}, \nn \\
c_{\bm k} &=&
- \frac{[{\bm U}_{\bm k}]^{*}_{3,1}
[{\bm U}_{\bm k}]^{*}_{4,2}
- [{\bm U}_{\bm k}]^{*}_{3,2}
[{\bm U}_{\bm k}]^{*}_{4,1}}
{[{\bm U}_{\bm k}]^{*}_{1,1}
[{\bm U}_{\bm k}]^{*}_{2,2}
- [{\bm U}_{\bm k}]^{*}_{1,2}
[{\bm U}_{\bm k}]^{*}_{2,1}}. \label{element}
\end{eqnarray}
Notice that $c_{\bm k} =
a_{\bm k} b^{\prime}_{\bm k} -
a^{\prime}_{\bm k} b_{\bm k}$.
This makes it possible to exponentiate the right hand side of
Eq.~(\ref{gs1}) as
\begin{eqnarray}
|{\rm g.s.} \rangle
\equiv \exp \Big[\sum_{{\bm k};k_y>0}
\big[{\bm t}_{\bm k}\big]_{\alpha\beta} \!\
f^{\dag}_{-{\bm k},\alpha}
f^{\dag}_{{\bm k},\beta}\Big] \!\ \big| 0 \big\rangle \label{gs2}
\end{eqnarray}
with
\begin{eqnarray}
{\bm t}_{\bm k}
&\equiv& \left[\begin{array}{cc}
a^{\prime}_{\bm k} & a_{\bm k} \\
b^{\prime}_{\bm k} & b_{\bm k} \\
\end{array}\right].
\end{eqnarray}

Equation~(\ref{gs2}) generally has a finite weight not only
on physical (i.e. spin) Hilbert space but also on those
fermionic states having either double occupancy on a
single site
or an empty site.
To obtain a variational many-body wavefunction in the
physical spin Hilbert space, we need to project out these unphysical
fermionic states, imposing `single-fermion condition'
on every site;
\begin{eqnarray}
\big|\Psi_{\bm \alpha} \big\rangle \equiv {\cal P}
\big|{\rm g.s.}\big\rangle, \label{projection}
\end{eqnarray}
where ${\cal P}$ stands for
the projection operator onto the
physical spin Hilbert space.
The projected BCS wavefunction
$|\Psi_{{\bm \alpha}}\rangle$
depends on the pairing
and hopping fields encoded in the
BdG Hamiltonian, such as $D$, $E$, $\chi$, $\eta$ and
$h_{\rm eff}$. The characteristic of the mean fields is
symbolically represented by
the subscript ${\bm \alpha}$.

Within the spin Hilbert space,
the wavefunction  is expressed by its inner-product
with an Ising spin configuration,
$|\{\sigma_{\bm j}\}\rangle=
\{\prod_{{\bm j}} f^{\dag}_{{\bm j},
\sigma_{\bm j}}\}|0\rangle$.
This product generally reduces to a Pfaffian,~\cite{pf-1,vmc,pf0}
\begin{eqnarray}
\langle \{\sigma_{\bm j}\}
|\Psi_{\alpha}\rangle
= {\rm Pf}\big[{\bm X}_{\alpha}(\{\sigma_{\bm j}\})
\big], \label{pfaffian}
\end{eqnarray}
where ${\bm X}_{\alpha}(\{\sigma_{\bm j}\})$ denotes
the $N \times N$ antisymmetric matrix given by~\cite{pf-1,pf0}
\begin{eqnarray}
\big[{\bm X}_{\alpha}(\{\sigma_{\bm j}\})
\big]_{{\bm j},{\bm l}}
&\equiv&
\big[ {\bm t}({\bm j},{\bm l})\big]_{\sigma_{\bm j},\sigma_{\bm l}}
- \big[ {\bm t}({\bm l},{\bm j})\big]_{\sigma_{\bm l},\sigma_{\bm j}}, \nn \\
\big[ {\bm t}({\bm j},{\bm l}) \big]_{\alpha,\beta}
&\equiv& \frac{1}{N} \sum_{{\bm k};k_y>0}
e^{i{\bm k}\cdot ({\bm j}-{\bm l})}
\big[{\bm t}_{\bm k}\big]_{\alpha,\beta}. \label{ft}
\end{eqnarray}
Note that the boundary condition for the momentum
$\bm k$ remains arbitrary in Eq.~(\ref{ft}).
To fix this arbitrariness, let us require that the
spin wavefunction given by Eq.~(\ref{pfaffian})
is an eigenstate of translations,
\begin{eqnarray}
\big\langle \{\sigma_{T_{a}({\bm j})} \} \big|\Psi_{\alpha}\big\rangle
 = e^{i\theta_{a}}   \big\langle \{\sigma_{\bm j} \}
\big|\Psi_{\alpha} \big\rangle, \label{trans}
\end{eqnarray}
where $T_{a}$ denotes the lattice translational operation
by ${\bm e}_a$, i.e.,
$T_{a}({\bm j})={\bm j}+{\bm e}_a$ $(a=x,y)$.
In fact, the preceding exact diagonalization
studies~\cite{note1}
suggest that the states with non-zero $Q$
vectors are unlikely realized in any intermediate
coupling regime of the present $J_1$--$J_2$
model, so that we impose the translational invariance
$(e^{i\theta_x},e^{i\theta_y})=(1,1)$ on
Eq.~(\ref{trans}). To satisfy this requirement, the
fermion's momenta in Eq.~(\ref{ft}) have only to
observe either the anti-periodic boundary condition
(APBC), i.e., $k_a=(2n_a-1)\pi /L_a$ with
$n_a=-L_a/2+1,\cdots,L_a/2$,
or the periodic boundary condition (PBC), i.e.,
$k_a=2n_a\pi /L_a$.
For the two-dimensional models, the trial
wavefunctions have four options. When both $k_x$
and $k_y$ satisfy the anti-periodic boundary
condition, the total momentum
carried by our trial spin wavefunction
is indeed at the $\Gamma$-point, i.e.
$(e^{i\theta_x},e^{i\theta_y})= (1,1)$; to see
this, one has only to relate
${\bm X}_{\alpha}(\{\sigma_{T_a({\bm j})}\})$
with  ${\bm X}_{\alpha}(\{\sigma_{\bm j}\})$
in terms of a certain elementary row/column
operation ${O}_{a}$,
\begin{eqnarray}
{\bm X}_{\alpha}(\{\sigma_{T_a({\bm j})}\})
= {O}^{T}_a\!\
{\bm X}_{\alpha}(\{\sigma_{\bm j}\})
{O}_a, \nn
\end{eqnarray}
where $O_a$ exchanges site
indices of $X_{\alpha}(\{\sigma_{\bm j}\}) $
according to the lattice translation
operator $T_a$.
Similarly, when $k_x$ satisfies
the periodic boundary condition
while $k_y$ does the anti-periodic boundary
condition or vice versa, the momenta carried by
the projected BCS wavefunction can be
shown to be
$(e^{i\theta_x},e^{i\theta_y})= ((-1)^{(L_y-1)L_x},1)$  or
$(e^{i\theta_x},e^{i\theta_y})= (1,(-1)^{(L_x-1)L_y})$,
respectively.
In what follows, we only consider
the systems with even length $L_x$ and $L_y$,
to impose the translational invariance of the total
wavefunction, i.e., $(e^{i\theta_x},e^{i\theta_y})= (1,1)$.

Notice also that, when both $k_x$ and $k_y$
observe the periodic boundary condition,
any of the projected BCS wavefunctions derived from
Eqs.~(\ref{j1},\ref{j2},\ref{j3}) cannot be
expressed in terms of a single Pfaffian.
This is roughly because, being either $d$-wave or
$p$-wave, all the pairing
fields in the Cooper channel always vanish at the four
time-reversal invariant momentums points
$(0,0)$, $(0,\pi)$, $(\pi,0)$ and $(\pi,\pi)$,
where a $4 \times 4$ BdG Hamiltonian reduces to a $2 \times 2$
Bloch Hamiltonian having
no anomalous part. As a result, the state-basis
representation of the projected BCS wavefunction
becomes relatively cumbersome.
In this paper, we study only those projected
BCS wavefunctions derived based on the other
three boundary conditions. Following the standard
literature,~\cite{wen,fradkin}
we name the projected BCS wavefunction
defined with the APBC in the both direction as
the `wavefunction in the $(\pi,\pi)$-topological
sector' and that with the PBC in one direction and
the APBC in the other as the `wavefunction in the
$(0,\pi)$ or $(\pi,0)$-topological sector.'

\subsection{Quantum spin number projection}
Our Hamiltonian has the global $SU(2)$ spin rotational
symmetry, while the trial wavefunctions constructed
from spin-triplet pairing states explicitly break this
continuous symmetry by hand.
Such a symmetry breaking is
supposed to occur only in the thermodynamic limit.
The ground-state wavefunction in a
finite-size system can be always identified as 
an eigenstate of the symmetry groups of the Hamiltonian.
Accordingly,
it is naturally expected 
that
the energy of the trial state will be further
improved, when the state being
projected onto the eigenspace of an appropriate
quantum number associated with
the spin-rotational symmetries. Thus,
we also consider as our trial state 
the projections of the
spin-triplet BCS wavefunctions with
{\it the quantum spin numbers}.

The spin projection operator which filters
out a state with the total spin $S=L$
and the $z$-component of the total spin $S_z=M$ has a form~\cite{MI,pf1}
\begin{align}
{\cal P}_{S_z=M}
{\cal P}_{S= L}
 \equiv &\frac{2L+1}{8\pi^2}
 \int^{2\pi}_{0} d\alpha \int^{\pi}_{0} d\beta \sin \beta
\int^{2\pi}_{0} d\gamma  \nn \\
&  \times 
\!\ P_{L}(\cos \beta)
\!\ \!\ e^{i\alpha (\hat{S}_z-M)}
e^{i\beta \hat{S}_y} e^{i\gamma \hat{S}_z}, \label{qsp}
\end{align}
where $P_L$ denotes the $L$-th Legendre
polynomial, and  ${\cal P}_{S=L}$ and ${\cal P}_{S_z=M}$ denote the projection operators
filtering out a state with the total spin $S=L$ and a state with
the $z$-component of the total spin $S_z=M$, respectively.

Combining this with Eqs.~(\ref{gs2})--(\ref{ft}),
we obtain the projected BCS wavefunction
with the quantum spin number projection as
\begin{align}
&
\langle \{\sigma_{\bm j}\}|
{\cal P}_{S_z=M}
{\cal P}_{S= L}
|\Psi_{\bm \alpha}\rangle
= \frac{2L+1}{4\pi} \times \nn \\
&
\int^{\pi}_{0} d\beta \sin \beta \int^{2\pi}_{0} d\gamma
\!\ P_L(\cos \beta) \!\ \!\
{\rm Pf}[{\bm X}_{\bm \alpha}(\{\sigma_{\bm j}\};\beta,\gamma)]
\label{rep}
\end{align}
under the condition
$\frac{1}{2}\sum_{\bm j} \sigma_{\bm j} = M$.
Here, the $N \times N$ antisymmetric matrix
${\bm X}_{\alpha}(\{\sigma_{\bm j}\};\beta,\gamma)$ 
is defined the same as in Eq.~(\ref{ft}) with the $2 \times 2$ matrix
${\bm t}({\bm j},{\bm l})$ 
being redefined in a rotated spin frame;
\begin{align}
\big[{\bm X}_{\alpha}(\{\sigma_{\bm j}\};\beta,\gamma)
\big]_{{\bm j},{\bm l}}
\equiv&
\big[ {\bm V}_{\beta,\gamma} {\bm t}({\bm j},{\bm l}){\bm V}^{-1}_{\beta,\gamma}
\big]_{\sigma_{\bm j},\sigma_{\bm l}} \nn \\
&- \big[ {\bm V}_{\beta,\gamma} {\bm t}({\bm l},{\bm j}) {\bm V}^{-1}_{\beta,\gamma}
\big]_{\sigma_{\bm l},\sigma_{\bm j}},
\end{align}
where
\begin{align}
{\bm V}_{\beta,\gamma} &\equiv \left[\begin{array}{cc}
\cos{\frac{\beta}{2}} &  - \sin{\frac{\beta}{2}} \\
\sin{\frac{\beta}{2}} & \cos{\frac{\beta}{2}} \\
\end{array}\right] 
\left[\begin{array}{cc}
e^{i\frac{\gamma}{2}} & 0 \\
0 &  e^{-i\frac{\gamma}{2}}  \\
\end{array}\right]. \nn
%
\end{align}

To integrate over $\beta$
and $\gamma$ numerically in Eq.~(\ref{rep}),
we employ the
Gauss-Legendre quadrature.
When projecting into the singlet space,
i.e. $S=0$, with the system size
$N=6\times 6 \sim 12 \times 12$,
we typically used $10 \sim 16$ mesh
points for the integration over $\beta$ and
$10 \sim 20$ mesh points for
that of $\gamma$.~\cite{as}

\section{energy optimization and energetics}
In this section, we optimize the energies
of the projected (i) $Z_2$ planar state,
(ii) $Z_2$ polar state and (iii) $SU(2)$ chiral
$p$-wave state and compare their minimized
energies with those of the ferromagnetic state
and the collinear antiferromagnetic state. Specifically,
we have numerically calculated the expectation
values of the energy for these projected BCS
wavefunctions, taking the quantum spin number
projection onto the subspace with either
$S_{z}=0$ or $S=0$;
\begin{eqnarray}
E^{S_z=0}_{\bm \alpha} =
\frac{\langle \Psi_{\bm \alpha} | \!\
{\cal P}_{S_z=0} \!\ H \!\ {\cal P}_{S_z=0}
\!\ |
\Psi_{\bm \alpha}\rangle}
{\langle \Psi_{\bm \alpha} |
{\cal P}_{S_z=0}
 |
\Psi_{\bm \alpha}\rangle}\label{Szt0}
\end{eqnarray}
and
\begin{eqnarray}
E^{S=0}_{\bm \alpha} =
\frac{\langle \Psi_{\bm \alpha} | \!\
{\cal P}_{S=0} \!\ H \!\ {\cal P}_{S=0}
\!\ |
\Psi_{\bm \alpha}\rangle}
{\langle \Psi_{\bm \alpha} |
{\cal P}_{S=0}
|
\Psi_{\bm \alpha}\rangle}. \label{St0}
\end{eqnarray}
We have further optimized these energies,
tuning the variational parameters ${\bm \alpha}$
encoded in the original BCS wavefunctions,
such as $D$, $E$, $\chi$, $\eta$ and $h_{\rm eff}$.
For this optimization,
we have employed the so-called stochastic
reconfiguration method.~\cite{SR,uf}

\subsection{Stochastic reconfiguration method}
Here we briefly review the stochastic
reconfiguration (SR) method.~\cite{uf,SR}
In this optimization method, a usual
steepest descent (SD) method is
modified in such a way that
information of the `quantum distance'
between wavefunctions is included.
The quantum distance is
chosen to be the square distance between
two normalized wavefunctions defined
in two different parameter points, say
${\bm \alpha}$ and
${\bm \alpha}+\delta {\bm \alpha}$,
in the form
\begin{equation}
\Delta^2_{\rm SR}
\equiv
\Big[\big \langle \overline{\Psi}_{{\bm \alpha}+\delta {\bm \alpha}}
\big| - \big \langle \overline{\Psi}_{\bm \alpha}\big|
\Big]\!\ \Big[\big | \overline{\Psi}_{{\bm \alpha}+\delta {\bm \alpha}}
\big\rangle  - \big | \overline{\Psi}_{\bm \alpha}\big\rangle
\Big]
\end{equation}
with
\begin{equation}
\big|\overline{\Psi}_{\bm \alpha} \big\rangle
= \big| \Psi_{\bm \alpha}\big\rangle
\{ \big\langle \Psi_{\bm \alpha}
\big|\Psi_{\bm \alpha}\big\rangle\}^{-\frac{1}{2}}
. \nn
\end{equation}
Regarding $|\delta \alpha|$ as a small
quantities, we can expand this quantum
distance in terms of $\delta {\bm \alpha}$,
\begin{equation}
\Delta^2_{\rm SR}= \sum_{j,m} \delta \alpha_j \delta \alpha_m
\left[{\bm S}_{\alpha}\right]_{j,m}
+ {\cal O}(\delta {\bm \alpha}^3),
\end{equation}
where the metric tensor $[{\bm S}_{\alpha}]$
is defined in the variational parameter space as
\begin{equation}
\left[ {\bm S}_{\alpha} \right]_{j,m}
 \equiv
\big \langle \partial_{\alpha_j}
\overline{\Psi}_{\bm \alpha}\big| \partial_{\alpha_m}
\overline{\Psi}_{\bm \alpha} \big\rangle + {\rm c.c.}.
\end{equation}
In the standard steepest descent (SD) method,
the variational parameters are changed
along the gradient of an
energy, 
$\delta \alpha_j =
\lambda \partial_{\alpha_j} E_{\bm \alpha}$
with
$E_{\bm \alpha} = \big\langle \overline{\Psi}_{\bm \alpha} \big|
{H} \big|\overline{\Psi}_{\bm \alpha} \big\rangle$ and
a small positive value $\lambda$. Meanwhile,
the SR method determines the optimal direction,
by minimizing the
energy {\it on the contour-(super)sphere of the equal
quantum distance}.
A variational principle with constraint
dictates that the optimal direction thus
defined is given by
$\delta \alpha_j = \lambda
\sum_m [{\bm S}^{-1}_{\alpha}]_{j,m}
\partial_{\alpha_m} E_{{\bm \alpha}}$.
It is empirically recognized that the
modification in terms of the metric tensor
$[{\bm S}_{\alpha}]$ substantially improves
the optimization efficiency, especially
when the tensor has a highly
non-flat structure in the variational
parameter space.~\cite{SR,uf}

The numerical evaluation of the metric tensor
and the gradient vector requires the summation
over all the Ising spin configurations
in the physical Hilbert space,
\begin{align}
&[{\bm S}_{\alpha}]_{m,n}
= \frac{1}{2}
\sum_{\{\sigma_{\bm j}\}}
\big(
{\cal O}^{*}_{m,\{\sigma_{\bm j}\}} {\cal O}_{n,\{\sigma_{\bm j}\}}
+ {\rm c.c.} \big)
w_{\{\sigma_{\bm j}\}} \nn \\
& \hspace{5mm}
-
\sum_{\{\sigma_{\bm j}\}}
{\rm Re} {\cal O}_{m,\{\sigma_{\bm j}\}}
w_{\{\sigma_{\bm j}\}}
\sum_{\{\sigma_{\bm m}\}}
{\rm Re}{\cal O}_{n,\{\sigma_{\bm m}\}}
w_{\{\sigma_{\bm m}\}}, \\
&\partial_{\alpha_m} E_{{\bm \alpha}}
= \sum_{\{\sigma_{\bm j}\}}
\big(
{\cal E}_{\{\sigma_{\bm j} \}} {\cal O}_{m,\{\sigma_{\bm j}\}}
 + {\rm c.c.} \big) \omega_{\{\sigma_{\bm j}\}} \nn \\
& \hspace{5mm}
- 2\sum_{\{\sigma_{\bm j}\}}
{\rm Re}{\cal O}_{m,\{\sigma_{\bm j}\}}
w_{\{\sigma_{\bm j}\}}
\sum_{\{\sigma_{\bm m}\}}
{\rm Re} {\cal E}_{\{\sigma_{\bm m}\}}
w_{\{\sigma_{\bm m}\}},
\end{align}
where
\begin{align}
w_{\{\sigma_{\bm j}\}} &=
\frac{|\langle \{\sigma_{\bm j}\}|\Psi_{\bm \alpha}\rangle |^2}
{\langle \Psi_{{\bm \alpha}} | \Psi_{{\bm \alpha}}\rangle},
\label{probability}\\
{\cal O}_{m,\{\sigma_{\bm j}\}} &=
\frac{\langle \{\sigma_{\bm j}\}|
\partial_{\alpha_m}
\Psi_{\bm \alpha} \rangle}
{\langle \{\sigma_{\bm j}\} | \Psi_{\bm \alpha}  \rangle},
\label{observableA}\\
{\cal E}_{\{\sigma_{\bm j}\}} &=
\frac{\big\langle \Psi_{\bm \alpha}\big|
H \big|\{\sigma_{\bm j}\} \big\rangle}{\big\langle
\Psi_{\bm \alpha} \big| \{\sigma_{\bm j}\}\big\rangle}.
\label{observableB}
\end{align}
The SR method replaces this extensive
summation by the statistical average
where $w_{\{\sigma_{\bm j}\}}$ is
regarded as a probability density of
the corresponding statistical ensemble.
Specifically, we numerically create a
Markov chain in which a binary
configuration $\{\sigma_{\bm j}\}$
is statistically generated with the
probability $w_{\{\sigma_{\bm j}\}}$.
In the statistical ensemble thus defined,
observables defined in
Eqs.~(\ref{observableA}-\ref{observableB})
are numerically evaluated;
\begin{eqnarray}
[{\bm S}_{\bm \alpha}]_{m,n}
&=& \frac{1}{2} \big(
\overline{{\cal O}_m \!\ {\cal O}^{*}_n}
 + {\rm c.c.}\big) - \frac{1}{4}
\big(\overline{{\cal O}_m} + {\rm c.c.}\big)
\big(\overline{{\cal O}_n} + {\rm c.c.} \big) , \nn \\
\partial_{{\bm \alpha}_m}{E}_{\bm \alpha}
&=& \big(
\overline{{\cal E} \!\ {\cal O}^{*}_m}
 + {\rm c.c.}\big) - \frac{1}{2}
\big(\overline{\cal E} + {\rm c.c.}\big)
\big(\overline{{\cal O}_m} + {\rm c.c.} \big).  \nn
\end{eqnarray}
To obtain the metric tensors and
gradient vector, we usually take
$1000\sim 4000$ samplings per site. For a set
of optimized variational parameters,
we evaluate the energy (sec.~IVB) and
the correlation function (sec.~VI),
where we typically use $10^5$
samplings. As for the projected BCS
wavefunction with the quantum spin
number projection, one has only to
replace $|\Psi_{\bm \alpha}\rangle$
in Eqs.~(\ref{probability})--(\ref{observableB})
by ${\cal P}_{S_z=M}
{\cal P}_{S = L}
|\Psi_{\bm \alpha}\rangle$.
Those who are interested in the actual
evaluation of these observables
can consult Refs.~\onlinecite{pf0}
and \onlinecite{pf1}.

\subsection{Energetics}
\begin{figure}
    \includegraphics[width=82.mm]{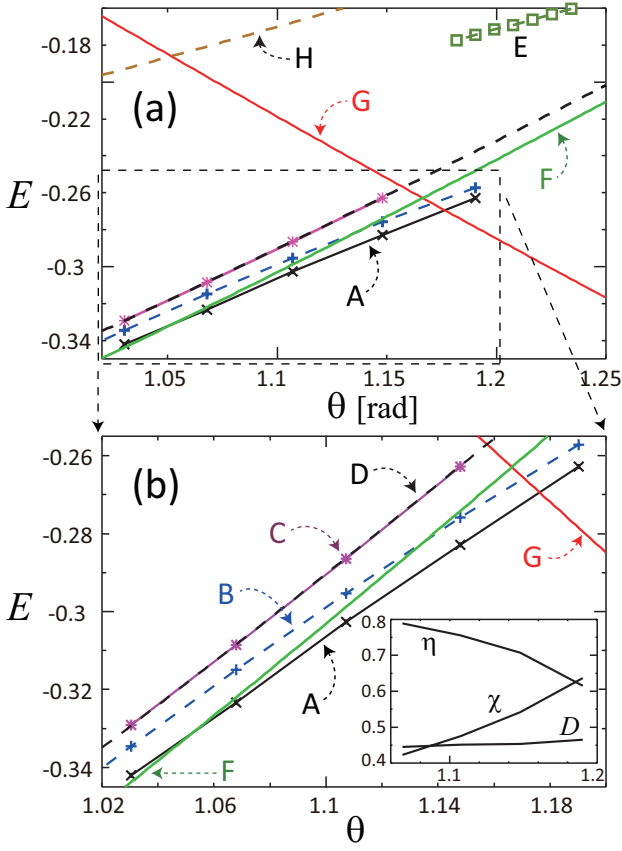}
\caption{(Color online) Energy comparison in
the square lattice $J_1$--$J_2$ model
as a function of
$\theta$ with $(J_1,J_2)=(-\sin \theta, \cos \theta)$.
(a): Optimized energies of the planar states [(A)
${\cal P}_{S=0}|\Psi_{\rm planar}\rangle$ and
(B) ${\cal P}_{S_z=0}|\Psi_{\rm planar} \rangle $],
the polar state [(C)
${\cal P}_{S_z=0}|\Psi_{\rm polar} \rangle$],
the $\pi$-flux state [(D) ${\cal P}_{S_z=0}
|\Psi_{\mbox{\tiny\rm $\pi$-flux}} \rangle $],
the $p$-wave chiral state
[(E) ${\cal P}_{S_z=0}|\Psi_{\rm chiral} \rangle$],
and the collinear antiferromagnetic state (F).
The exact energies of the ferromagnetic state (G) and
the isolated dimer state (H) are also shown.
The projected state ${\cal P}_{S=0}|\Psi_{\rm planar}\rangle$
is calculated in $8\times 8$ spin system and
${\cal P}_{S_z=0}|\Psi_{\rm planar} \rangle $ is in
$10\times 10$ spin system.
(b): A part of figure (a) is enlarged.
(inset): Optimized parameter values of $D$,
$\chi$, and $\eta$ in
${\cal P}_{S=0}|\Psi_{\rm planar} \rangle $
as a function of $\theta$. These values are
rescaled, such that $\sqrt{D^2+\chi^2+\eta^2}=1$.
The optimal values of the
excitonic spin-triplet pairing field $E$ and
the effective Zeeman field $h_{\rm eff}$ are
negligibly small.
}
\label{fig:ene-the}
\end{figure}
Figure~\ref{fig:ene-the} shows the
optimized energies of
the projected $Z_2$ planar states (both
${\cal P}_{S_z=0}|\Psi_{\rm planar}\rangle$ and
${\cal P}_{S=0}|\Psi_{\rm planar}\rangle$),
the projected $Z_2$ polar state
${\cal P}_{S_z=0}|\Psi_{\rm polar}\rangle$,
and the projected $SU(2)$ chiral $p$-wave state
${\cal P}_{S_z=0}|\Psi_{\rm chiral}\rangle$.
We also compare these optimized energies
with the exact energy of the fully polarized ferromagnetic
state $E_{\rm ferro}=-0.5(|J_1|-J_2)$,
and the variationally optimized energies
of the collinear antiferromagnetic state~\cite{LDA}
$E_{\rm CAF}= -0.6682 J_2$
and the decoupled double $\pi$-flux
state~\cite{Gros} $E_{\mbox{\tiny\rm $\pi$-flux}}
= -0.64 J_2$.
The collinear antiferromagnetic state we considered
in this paper is composed by
two `decoupled' N\'eel-ordered states
\begin{eqnarray}
|\Psi_{\rm CAF} \rangle =
|\Psi_{\rm Neel}\rangle_{\rm A}
|\Psi_{\rm Neel}\rangle_{\rm B}, \label{Neel}
\end{eqnarray}
each of
which is defined on a non-frustrated square
sublattice coupled with $J_2$ bonds, say A-sublattice
or B-sublattice.
For $|\Psi_{\rm Neel}\rangle$,
we employed the
variational wavefunction numerically derived
by Liang et.al.,~\cite{LDA} which was maximally
optimized on the antiferromagnetic
square lattice in terms of the staggered magnetic
moment, singlet pairing fields, and
the Jastrow factor. 
Note that the expectation value of the ferromagnetic
exchange interaction always vanishes in
Eq.~(\ref{Neel}), though the spin-triplet
pairing fields connecting the
two sublattices could possibly decrease
the energy in general.
In spite of this, however, the energy
of Eq.~(\ref{Neel}) achieves about
94.5$\%$ of the exact ground
state energy in 36 spin cluster
at $|J_1| = J_2$, whereas it achieves about $96.3\%$
at $J_1 = 0$. We thus regard that,
even in the presence of considerable
$J_1$,  Eq.~(\ref{Neel}) still gives an
appropriate energetics for the collinear
antiferromagnetic state of
the $J_1$--$J_2$ model.

\begin{figure}
    \includegraphics[width=70mm]{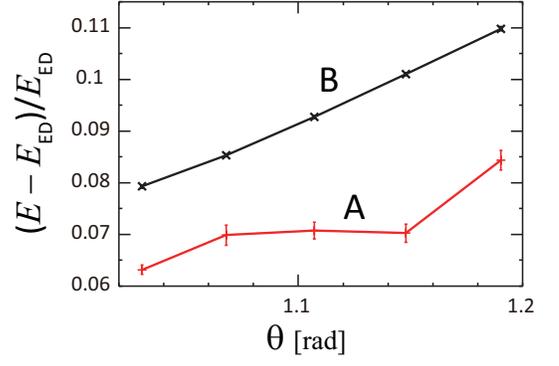}
\caption{(Color online)
Comparison between the optimized energies
of the projected planar states (A) ${\cal P}_{S=0}|\Psi_{\rm planar}\rangle$
and (B) ${\cal P}_{S_z=0}|\Psi_{\rm planar}\rangle$,
and the ground state energy $E_{\rm ED}$ obtained
by exact diagonalization,
in the square lattice $J_1$--$J_2$ model,
as a function of $\theta=\tan^{-1}(-J_1/J_2)$. The system size
is $6\times 6$. }
\label{fig:cf}
\end{figure}

The energy comparison in Fig.~\ref{fig:ene-the}
shows that the projected $Z_2$ planar
state has the lowest energy in
a finite range of the intermediate coupling
regime,
$0.417 |J_1| \alt J_{2} \alt 0.57 |J_1|$, whereas
the ferromagnetic state is the most stable in the strong
$J_1$ regime, $J_{2} \alt 0.417 |J_1|$,
and the collinear antiferromagnetic state is in the strong $J_2$ regime,
$0.57 |J_1| \alt J_{2}$.
The optimal energy of the
planar state projected onto the $S_z=0$ sector in the $6\times 6$ system
achieves roughly $92\% \sim 89\%$ of the exact ground state
energy obtained by the numerical diagonalization
with the same system size.  The energy becomes further
decreased by $2\% \sim 3\%$, when the wavefunction is
projected onto the $S=0$ space (see Fig.~\ref{fig:cf}).

Figure \ref{fig:ene-the} suggests that,
contrary to 
the mean-field analysis,
the projected chiral $p$-wave state
hardly realizes in any of the intermediate
coupling regime of the $J_1$--$J_2$ model,
at least when the system size is $2n
\times 2n$ ($n=3,4,5\cdots$). Moreover,
the estimated energy  of
${\cal P}_{S_z=0}|\Psi_{\rm chiral}\rangle$
is already $30\%$
higher than those of
the ferromagnetic state
and the projected $Z_2$ planar state,
so that the situation is unlikely
reversed, even when the wavefunction is
further projected into the singlet space, i.e.
${\cal P}_{S=0} |\Psi_{\rm chiral}\rangle$.
Figure~\ref{fig:ene-the} also indicates that
the $Z_2$ polar state is almost energetically
degenerate with the double $\pi$-flux state,
which indicates that the spin-triplet pairing
does not lower the energy efficiently in
this state. Indeed,
we observed that the optimized value of the
triplet pairing field in the polar state is less
than $10\%$ of the root square sum of all the
variational parameters.

\begin{figure}
    \includegraphics[width=85mm]{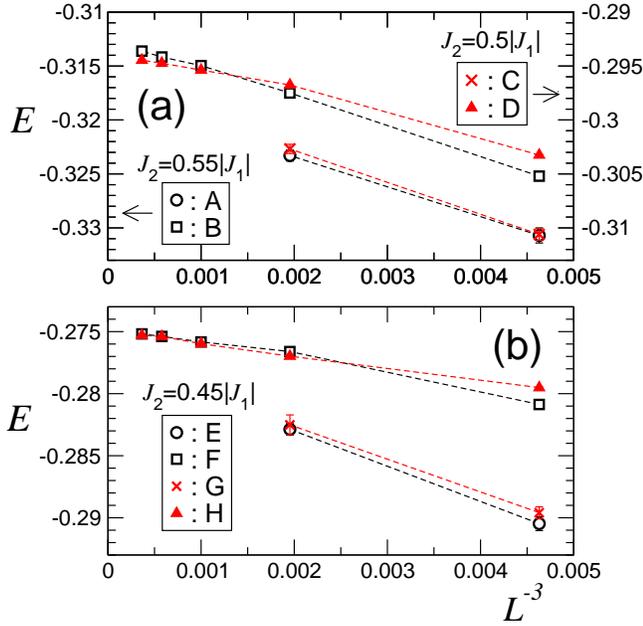}
\caption{(Color online) Size dependence of the optimized energies per site
for the projected planar states in the $L\times L$ square lattice
$J_1$--$J_2$ model
with $J_2=0.55 |J_1|$ (a: black dashed lines),
$J_2=0.5 |J_1|$ (a: red dashed lines), and
$J_2=0.45 |J_1|$ (b). The energy unit is taken to
be $\sqrt{J^2_1 + J^2_2}$.
The projected states are taken as
${\cal P}_{S_z=0}|\Psi_{\rm planar}\rangle$ ($L=6\sim 14$) for the data
(B, D, F, H)
and as ${\cal P}_{S=0}|\Psi_{\rm planar}\rangle$
($L=6,8$) for the data (A, C, E, G).
The horizontal axis is taken as $L^{-3}$.
For ${\cal P}_{S_z=0}|\Psi_{\rm planar}\rangle$
with $L=14$, we used the same variational parameters as those
of ${\cal P}_{S_z=0}|\Psi_{\rm planar}\rangle$ with
$L=12$. For ${\cal P}_{S=0}|\Psi_{\rm planar}\rangle$
with $L=8$, we used the same variational parameters as those
of ${\cal P}_{S=0}|\Psi_{\rm planar}\rangle$ with
$L=6$.
The data G and H are calculated in the $(0,\pi)$-topological sector
and the others are in the $(\pi,\pi)$-topological
sector.}
\label{fig:fi-si-sc}
\end{figure}

We also show the system size dependence of the optimized energy
per site for
the projected planar states in Fig.~\ref{fig:fi-si-sc}.
The data for the state ${\cal P}_{S_z=0}|\Psi_{\rm planar}\rangle$
indicates that the energy has a finite-size correction in the form
$E(L)/L^2=\epsilon_0 - c/L^3$.
This observation is consistent with the
existence of gapless
Goldstone modes in the $Z_2$ planar state.
Using the random phase approximation,
we can show that the low energy dispersions
of these gapless modes are always linear in
the momentum,~\cite{sym}
which suggests that the finite-size correction to the ground state energy
per site decays in the form
$ - c/L^3$, the same as that of
the two-dimensional antiferromagnetic
Heisenberg model.~\cite{huse}

\section{$d$-wave spin-nematic character of
projected $Z_2$ planar states}
In this section, we show that %
all the projected spin-triplet RVB states
derived from Eqs.~(\ref{j1},\ref{j2},\ref{j3})
generally have  the `spin-nematic' properties;
ordering of quadrupole moments without
{\it spontaneous ordering} of magnetic dipole
moments. In particular, we argue that the projected
$Z_2$ planar state 
has a `$d$-wave' spin-nematic character,
or an `$d$-wave' quadrupolar order,
which is consistent with the nature
of the spin nematic phase suggested by
the exact diagonalization study \cite{sms}.
All symmetries of the Anderson tower
of spin nematic states are clarified
from the decomposition of the $Z_2$
planar state.

To see the `spin-nematic' character,
notice first that
all the mean-field states discussed in Sec.~\ref{sec:MF} are
invariant under the spin $\pi$-rotation
about the 3-axis. Namely, spin-triplet
$d$-vectors in these states are always lying in a plane
perpendicular to the field (see Fig.~1),
so that the spin $\pi$-rotation around the field
changes the sign of the triplet pairing
fields on the nearest-neighbor ferromagnetic bonds, while leaves
intact the singlet pairing fields on the
next-nearest-neighbor antiferromagnetic bonds.
This sign change can be
readily set off by the staggered gauge transformation
$f^{\dag}_j \rightarrow f^{\dag}_j \!\ (-1)^{j_x + j_y}$.
The whole unitary transformation is expressed as
\begin{align}
U=&\exp\left[i\pi\sum_{j,\sigma} (j_x+j_y)f^\dagger_{j,\sigma} f_{j,\sigma}\right] \nonumber\\
&\ \ \ \times\exp\left[i\frac{\pi}{2}\sum_{j} f^\dagger_{j,\alpha} [\sigma_3]_{\alpha\beta} f_{j,\beta}\right].
\end{align}
Since the mean-field Hamiltonian is invariant
under this transformation, the transformed
state $U|\rm{g.s.}\rangle$ is energetically
degenerate with the original ground state
$|\rm{g.s.}\rangle$.
On the one hand, being a vacuum state of
the Bogoliubov particle, the ground state
should be unique, which leads
to $U|\rm{g.s.}\rangle=e^{i\theta}|\rm{g.s.}\rangle$.
Thereby, the projected BCS wavefunction
generally satisfies the following relation
\begin{eqnarray}
\langle \{\sigma_{\bm j}\}| \Psi_{\bm \alpha} \rangle
=
\langle \{\sigma_{\bm j}\}| {\cal P}|{\rm g.s.} \rangle
=
e^{-i\theta}
\langle \{\sigma_{\bm j}\}| {\cal P} U |\rm{g.s.}\rangle.
\end{eqnarray}
Since the unitary operator $U$ commutes
with the projection ${\cal P}$, the right
hand side can be further written as follows
with the eigenvalues
$S_z \equiv \frac{1}{2}\sum_{\bm j}\sigma_{\bm j}$,
\begin{eqnarray}
\langle \{\sigma_{\bm j}\}|\Psi_{\bm \alpha}\rangle
= e^{-i\theta} e^{i\pi S_z}
\langle \{\sigma_{\bm j}\} |\Psi_{\bm \alpha} \rangle. \label{pi-rot4}
\end{eqnarray}
One can easily fix the $U(1)$ phase factor,
evaluating the product between the
projected BCS wavefunctions and a fully polarized
state $|\{\sigma_{\bm j}=1\}\rangle
\equiv \{\prod_{{\bm j}}
f^{\dag}_{{\bm j},\uparrow}\}|0\rangle$,
\begin{eqnarray}
\ \
|\langle \{\sigma_{\bm j}=1\}|
\Psi_{\bm \alpha}\rangle|^2
=
\prod_{{\bm k};k_y>0} \big|a^{\prime}_{\bm k}\big|^2, \nn
\end{eqnarray}
where $a^{\prime}_{\bm k}$ is defined in
Eq.~(\ref{element}). Supposing
that $a^{\prime}_{\bm k}$ is non-vanishing
at any discretized momentum point at $k_y>0$,
which actually holds true for the
$Z_2$ planar state in the zero field case,
i.e. Eq.~(\ref{j0}), the projected BCS wavefunction
has a finite weight in the eigenspace
of $S_z=\frac{N}{2}$,
$|\langle \{\sigma_{\bm j}=1\}|
\Psi_{\bm \alpha}\rangle| \ne 0$.
To make this observation compatible with
Eq.~(\ref{pi-rot4}), the phase factor
must take a form,
\begin{eqnarray}
e^{-i\theta}=(-1)^{\frac{N}{2}}.
\label{phase}
\end{eqnarray}
In general, we can prove
Eqs.~(\ref{pi-rot4}-\ref{phase})
more directly, only by imposing
the spin-$\pi$ rotational symmetry onto the
eigenvectors of a given Bogoliubov
Hamiltonian.

Equation~(\ref{pi-rot4}) 
guarantees the `spin-nematic' character
of the spin-triplet RVB states. Namely, when
combined with Eq.~(\ref{phase}), this identity
requires that the wavefunctions have a finite
weight only in the subspace with an even-integer
$S_{z}$ for $N=4l$ $(l=1,2,\cdots)$ spin systems,
whereas only in the subspace with an odd-integer
$S_z$ for $N=4l+2$ $(l=0,1,\cdots)$ spin systems.
Thus, the transverse local magnetization always
vanishes in these projected spin-triplet RVB states,
\begin{equation}
\langle \Psi_{\bm \alpha}|S_{{\bm j},\pm}|
\Psi_{\bm \alpha}\rangle = 0,
\end{equation}
while the spin quadrupole moments
in the transverse plane are allowed to have a finite value,
\begin{eqnarray}
\langle \Psi_{\bm \alpha} | S_{{\bm j},+}
S_{{\bm m},+} | \Psi_{\bm \alpha}\rangle \equiv
f\big({\bm j}-{\bm m}\big) \ne  0, \label{transM}
\end{eqnarray}%
where
$S_{{\bm j},+} S_{{\bm m},+}$ relates
to the spin nematic operators in the form\cite{sms}
$S_{{\bm j},+} S_{{\bm m},+}=(K^{11}_{{\bm j},{\bm m}}
-K^{22}_{{\bm j},{\bm m}})
+2iK^{12}_{{\bm j},{\bm m}}$.

As for the $Z_2$ planar state derived from
Eqs.~(\ref{j0},\ref{j1}), this quadrupole moments
obey the $d$-wave spatial configuration,
\begin{eqnarray}
f\big(R_{\frac{\pi}{2}}({\bm j}- {\bm m})\big)
= - f\big({\bm j}-{\bm m}\big), \label{d-wave}
\end{eqnarray}
where $R_{\theta}$ denotes the space $\theta$-rotation
around the axis perpendicular to the square-lattice plane.
This $d$-wave nature comes from the fact that
the planar state is invariant under
the space $\frac{\pi}{2}$-rotation 
accompanied by the spin $\frac{\pi}{2}$-rotation
around the $3$-axis (around the field)
and the gauge transformation
$f^{\dag}_{{\bm j},\sigma} \rightarrow
i (-1)^{j_x+j_y} f^{\dag}_{{\bm j},\sigma}$.
Namely, the state is
invariant under the following unitary
transformation,
\begin{align}
U^\prime=&
\exp\left[i\pi\sum_{j,\sigma} \left(j_x+j_y+\frac12\right)f^\dagger_{j,\sigma} f_{j,\sigma}\right] \nonumber\\
&\ \ \ \ \times\exp\left[i\frac{\pi}{4}\sum_{j} f^\dagger_{j,\alpha} [\sigma_3]_{\alpha\beta} f_{j,\beta}\right]
R_{\frac{\pi}{2}}.
\end{align}
Utilizing this symmetry in the
same way as we did for the spin $\pi$-rotation
above, one can derive the following identity
for the projected $Z_2$ planar state,
\begin{eqnarray}
\langle \{\sigma_{\bm j}\}|\Psi_{\rm planar}\rangle
= (-1)^{\frac{N}{4}} \!\ e^{i\frac{\pi}{2} S_z }
\langle \{\sigma_{R_{\frac{\pi}{2}}(\bm j)}\}
|\Psi_{\rm planar}\rangle.
\label{piov2-rot}
\end{eqnarray}
To obtain the $d$-wave character from this identity,
expand the projected planar state into each
$S_z$-subspace. Under Eq.~(\ref{pi-rot4}),
it takes a form,
\begin{eqnarray}
&& |\Psi_{\rm planar} \rangle
 = \cdots
 +  {\cal P}_{S_z=-2} |\Psi_{\rm planar} \rangle
+ {\cal P}_{S_z=0}
|\Psi_{\rm planar} \rangle  \nn \\
&&\ \ \  \ \ + \!\
{\cal P}_{S_z=2} |\Psi_{\rm planar} \rangle
+ {\cal P}_{S_z=4} |\Psi_{\rm planar} \rangle
+ \cdots, \label{expan}
\end{eqnarray}
for $N=4l$ ($l=1,2,\cdots$) spins.
Among this set of states, the
quadrupole operator
connects only those two states whose
$S_z$ differ by 2,
\begin{eqnarray}
&& f(\bm j- \bm m) = \sum_n \nn \\
&& \ \langle \Psi_{\rm planar} |
{\cal P}_{S_z = 2n+2} \!\
 S_{{\bm j},+} S_{{\bm m},+}
\!\ {\cal P}_{S_z =2n} |
\Psi_{\rm planar} \rangle. \nn
\end{eqnarray}
From Eq.~(\ref{piov2-rot}),
one of these two states is always even 
under the space $\frac{\pi}{2}$-rotation,
whereas the other is odd.  This clearly
assigns the $d$-wave spatial configuration
of the quadrupole moments,
i.e. Eq.~(\ref{d-wave}).

The arguments so far also
suggest 
how to construct a trial wavefunction for
the so-called Anderson's
tower of states (or quasi-degenerate joint states) of the
symmetry breaking $d$-wave spin nematic order
under the field. For $N=4l$
spin clusters,
${\cal P}_{S_z=2n}|\Psi_{\rm planar}\rangle$
mimic the quasi-degenerate joint states (QDJS)
of the spin nematic ordered phase, whose
representation under the point group
symmetry operators  are listed in Table.~I.
These representations are actually consistent
with those of the Bose-Einstein condensate phase
of a two-magnon bound state near the
saturation field.~\cite{sms}

\begin{table}
\caption{Indices of the projected $Z_2$ planar states
under the point group symmetries of the
square lattice, where
$N$ (multiples of $4$) denotes the total
number of the lattice points.
}
\begin{ruledtabular}
\begin{tabular}{cc}
point group symmetries &
${\cal P}_{S_z=2n} 
|\Psi_{\rm planar}\rangle$  \\  \hline
time-reversal (for $n=0$ and zero-field case) & 1 \\
$\frac{\pi}{2}$-rotation within
the lattice & $(-1)^{\frac{N}{4} + n}$ \\
mirror with respect to the $x$-link & $1$ \\
translations & $1$ \\
\end{tabular}
\label{table1}
\end{ruledtabular}
\end{table}

The projected $Z_2$ planar state in the zero-field case
is time-reversal symmetric; the wavefunction derived
from Eq.~(\ref{j0}) preserves
the following symmetry property,
\begin{align}
&
{\Pf} \big[{\bm X}_{\rm planar}(\{\sigma_{\bm n}\})\big]^{*} \nonumber\\
&= (-1)^{\frac{N}{2}}
\!\ \Big\{\prod_{\bm j} (-1)^{\sigma_{\bm j}} \Big\} \!\
{\Pf} \big[{\bm X}_{\rm planar}(\{-\sigma_{\bm n}\})\big].
\label{time-reverse}
\end{align}
To see this relation, notice first that the
time-reversal operation changes the sign of
the triplet Cooper pairing fields in Eq.~(\ref{j0}),
while does not affect 
the singlet pairing fields.
Such a sign change can be
readily compensated by the previous
staggered gauge transformation,
$f^{\dag}_j \rightarrow
f^{\dag}_j \!\ (-1)^{j_x + j_y}$,
which imposes the following relation onto
the BCS gap functions:
\begin{eqnarray}
\big[{\bm t}_{\bm k}\big]^{*} = {\bm \sigma}_2
\big[{\bm t}_{-{\bm k}+(\pi,\pi)}\big] {\bm \sigma}_2. \nn
\end{eqnarray}
Or equivalently,
\begin{eqnarray}
&&
\big[{\bm X}_{\rm planar}(\{\sigma_{\bm n}\})
\big]^{*}_{\bm j\bm m}
= (-1)^{j_x+j_y} (-1)^{m_x+m_y} \nn \\
&& \hspace{2cm} \times (-1)^{\sigma_{\bm j}} (-1)^{\sigma_{\bm m}}
\big[{\bm X}_{\rm planar}(\{-{\sigma}_{\bm n}\})
\big]_{\bm j \bm m}. \nn
\end{eqnarray}
Noting that ${\rm Pf}[{\bm O}^T {\bm A}{\bm O}]=
\det {\bm O} \!\ {\rm Pf}[{\bm A}]$, one immediately
obtain Eq.~(\ref{time-reverse}). 
This equation especially means
that, as far as $| \Psi_{\rm planar}\rangle$
is constructed from Eq.~(\ref{j0}),
both ${\cal P}_{S_z=0}| \Psi_{\rm planar}\rangle$
and ${\cal P}_{S=0} | \Psi_{\rm planar}\rangle$
are even under the time reversal operation
for $N=4l$ spin systems, which is also consistent
with the nature of the spin nematic phase
suggested by the previous exact diagonalization
studies.~\cite{sms}

\section{static correlation functions}
\begin{figure}
    \includegraphics[width=86mm]{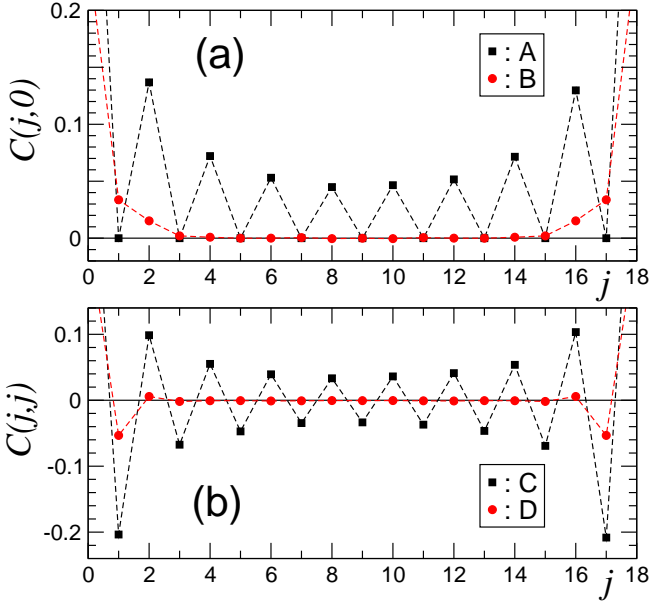}
\caption{(Color online)
Spin correlation functions of the transverse component $C_{\pm}({\bm j})$
(A and C)
and the longitudinal one $C_{zz}({\bm j})$ (B and D)
in the projected $Z_2$ planar state
${\cal P}_{S_z=0}|\Psi_{\rm planar}\rangle$ along (a) the
$x$-direction (1,0) and (b) the diagonal direction (1,1).
[A: $C_{\pm}(j,0)$, B : $C_{zz}(j,0)$,
C: $C_{\pm}(j,j)$ and D: $C_{zz}(j,j)$.]
The projected planar state was obtained in the $J_1$--$J_2$ model with
$J_1=-1$ and $J_2=0.45$
in $18 \times 18$ spin cluster, which takes the
variational parameters as
$(D,\chi,\eta)=(0.40,0.57,0.72)$.
The error-bar is smaller
than the symbol.}
\label{fig:cr-np}
\end{figure}
Based on energy comparison and symmetry arguments,
we have argued so far that the projected $Z_2$ planar state is
likely to be realized in the square lattice $S=1/2$ $J_1$--$J_2$
frustrated ferromagnetic model in the
intermediate coupling range $0.417 |J_1| \alt  J_2 \alt 0.57 |J_1|$.
To give a direct physical characterization to
this intermediate phase, we discuss in this
section the static correlation functions
calculated with respect to
${\cal P}_{S_z=0}|\Psi_{\rm planar}\rangle$
and ${\cal P}_{S=0}|\Psi_{\rm planar}\rangle$.
From the energetics, 
it is clear that
${\cal P}_{S=0}|\Psi_{\rm planar}\rangle$
is closer to the symmetric ground state of finite
spin clusters than the other in the intermediate phase.
On the other hand, the correlation
function of ${\cal P}_{S_z=0}|\Psi_{\rm planar}\rangle$
offers a feature of spin-rotational symmetry
broken spin nematic state.

Let us begin with 
the spin correlation function
calculated with respect to
${\cal P}_{S_z=0}|\Psi_{\rm planar}\rangle$.
In Fig.~\ref{fig:cr-np},
we show the characteristic behavior
of the transverse component of the spin correlation
function (labeled as `A' and `C'),
\begin{align}
C_{+-}({\bm j}-{\bm m}) &= \frac12
\big\langle \Psi_{\rm planar} \big| {\cal P}_{S_z=0}
\!\ \big\{\hat{S}_{{\bm j},+} \hat{S}_{{\bm m},-} \nn \\
& \hspace{1cm} +
\hat{S}_{{\bm j},-} \hat{S}_{{\bm m},+}  \big\} \!\
{\cal P}_{S_z=0} \big| \Psi_{\rm planar} \big\rangle,  \nn
\end{align}
and the longitudinal one (labeled as `B' and `D'),
\begin{align}
C_{zz}({\bm j}-{\bm m}) &=
\big\langle \Psi_{\rm planar} \big| {\cal P}_{S_z=0}
\!\ \hat{S}_{{\bm j},z} \hat{S}_{{\bm m},z} \!\
{\cal P}_{S_z=0} \big| \Psi_{\rm planar} \big\rangle.  \nn
\end{align}
Observing them, notice first that the transverse
component of the spin in the $A$-sublattice in which $j_x+j_y$ is even has no
correlation at all with those in the $B$-sublattice  in which $j_x+j_y$ is odd.
More generally,
this feature holds true for any (projected)
$Z_2$ planar state derived from Eq.~(\ref{j0}) or (\ref{j1}),
i.e.,
\begin{eqnarray}
\big\langle \Psi_{\rm planar} \big|
\big\{\hat{S}_{{\bm j},+} \hat{S}_{{\bm m},-} +
\hat{S}_{{\bm j},-} \hat{S}_{{\bm m},+}  \big\}
\big| \Psi_{\rm planar} \big\rangle
= 0
\label{elephantvanishes}
\end{eqnarray}
for $\forall\!\ {\bm j}\in A$
and $\forall\!\ {\bm m} \in B$.
Equation~(\ref{elephantvanishes}) can be
understood from the symmetry argument.
Suppose that the spin $\theta$-rotation
around the $z$-axis
is applied onto
all the spins in the $A$-sublattice, while the spin
$-\theta$-rotation is in
the $B$-sublattice. The mean-field Hamiltonian for
the $Z_2$ planar state 
is invariant under this continuous
transformation, so that the staggered magnetization
$S_{A,z}-S_{B,z} =\frac12
\sum_{{\bm j}\in A} \sigma_{\bm j} -
\frac12\sum_{{\bm j}\in B} \sigma_{\bm j}$ is a
conserved quantity. Applying this staggered
spin rotation
to the projected $Z_2$ planar states,
we obtain
\begin{equation}
\big\langle \{\sigma_{\bm j}\}\big|\Psi_{\rm planar}
\big\rangle = e^{i\theta (S_{A,z}-S_{B,z})}
\big\langle \{\sigma_{\bm j}\}\big|\Psi_{\rm planar}
\big\rangle \label{local1}
\end{equation}
for any $\theta$.
Equivalently, we have
\begin{eqnarray}
\big\langle \{\sigma_{\bm j}\}\big|\Psi_{\rm planar}
\big\rangle = \delta_{S_{A,z},S_{B,z}} \!\
g(\{\sigma_{\bm j}\}),
\end{eqnarray}
which immediately leads to
Eq.~(\ref{elephantvanishes}).

Equation~(\ref{elephantvanishes}) suggests that,
as for the transverse component of the spin
correlation function,
the next-nearest-neighbor antiferromagnetic
interaction dominates over the competing nearest-neighbor
ferromagnetic interaction.
In fact, the transverse spin exhibits a strong
antiferromagnetic correlation on each of the
`unfrustrated'  sublattice (square lattice with $J_2$ bonds), where the
function decays nearly in a power-law. 
(see `A' and `C' in Fig.~\ref{fig:cr-np}).
On the one hand, the longitudinal component
always has a
ferromagnetic correlation between the nearest
neighbor spins and an antiferromagnetic correlation between the 2nd neighbor
spins.
Thus, 
the spin-frustration among the ferromagnetic bonds
and antiferromagnetic ones
effectively suppresses the overall amplitude of
the longitudinal correlation function
(see `B' and `D' in Fig.~\ref{fig:cr-np}).
Indeed, $C_{zz}({\bm j})$ decays quite rapidly and
falls below $10^{-2}$ when
the spins are spatially separated by more than
two sites, $|{\bm j}|>3$.

\begin{figure}
    \includegraphics[width=86mm]{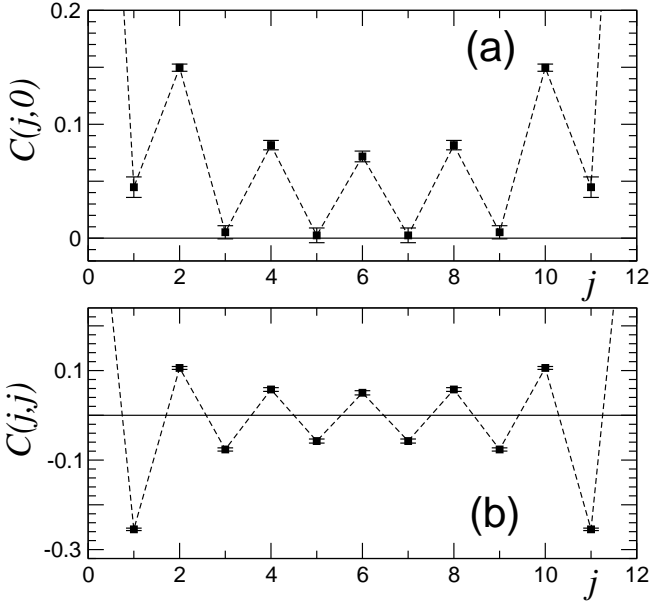}
\caption{Spin correlation function $C({\bm j})$ in the spin-singlet state
${\cal P}_{S=0}|\Psi_{\rm planar}\rangle$ along the
$x$-direction (a) and along the diagonal direction (b);
[(a) $C(j,0)$ and (b) $C(j,j)$].
The projected planar state was obtained for the $J_1$--$J_2$ model with $J_2=0.45 |J_1|$
in $12 \times 12$ spin cluster, which takes the
variational parameters as
$(D,\chi,\eta)=(0.45,0.55,0.70)$.}
\label{fig:cr-p}
\end{figure}

\begin{figure}
    \includegraphics[width=82mm]{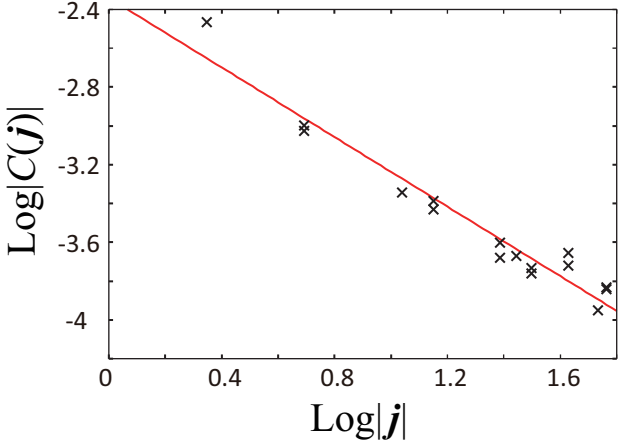}
\caption{(Color online)
A log-log plot of the correlation function $C({\bm j})$
in the spin-singlet state
${\cal P}_{S=0}|\Psi_{\rm planar}\rangle$
with the same parameter set as used in Fig.~\ref{fig:cr-p}.
We employ only those points with $\max(j_x,j_y)\le 6$
and $j_x+j_y={\rm even}$. 
The red solid line
indicates the slope of
$C({\bm j})\sim (-1)^{j_x}|{\bm j}|^{-0.9}$.}
\label{fig:pw-p}
\end{figure}

\begin{figure}
    \includegraphics[width=82mm]{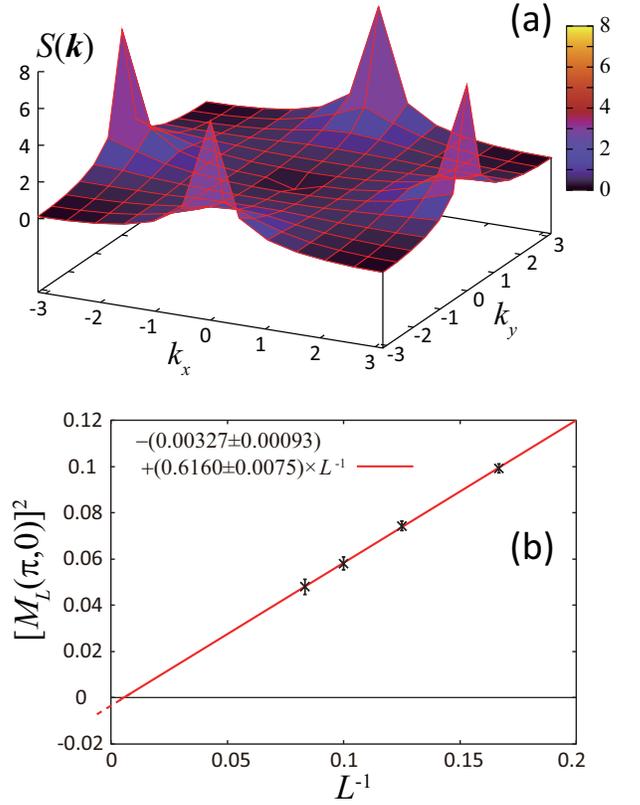}
\caption{(Color online)
(a)
Static spin structure factor $S({\bm k})$ calculated
from ${\cal P}_{S=0}
|\Psi_{\rm planar}\rangle$ with the same parameter set as used in Fig.~\ref{fig:cr-p}.
The momentum vector ${\bm k}$ ranges
over the 1st Brillouin zone,
$[-\pi,\pi]\times [-\pi,\pi]$.
(b) Finite size scaling of
$[M_{L}(\pi,0)]^2 = \frac{1}{N+2} S(\pi,0)$
as a function of the linear dimension
of the system size, indicating that
$M_{L}(\pi,0)$ converges
to zero in the thermodynamic limit within the statistical error of
the Monte Carlo estimation.}
\label{fig:FT}
\end{figure}

When the wavefunction is projected
onto the  spin singlet space,
the static spin correlation function
\begin{equation}
C({\bm j}-{\bm m}) = \big\langle \Psi_{\rm planar} \big|
{\cal P}_{S=0} \!\
\hat{\bm S}_{\bm j} \cdot \hat{\bm S}_{\bm m}
\!\ {\cal P}_{S=0}
\big| \Psi_{\rm planar} \big\rangle \label{invariant}
\end{equation}
becomes spin-rotational
invariant,
which `interpolate' between $C_{+-}({\bm j})$ and
$C_{zz}({\bm j})$ described above.
Namely, as shown in Fig.~\ref{fig:cr-p}, the correlation between
the spins in the $A$-sublattice and those in
the $B$-sublattice are either extremely short-range
or almost quenched, while
the correlation within each sublattice exhibits
an antiferromagnetic `quasi-long-ranged' power law decay,
which is fitted as 
$C({\bm j})\sim (-1)^{j_x}
|{\bm j}|^{-\eta}$ with $\eta = 0.9\sim 1.0 $ 
for $j_x+j_y={\rm even}$
(see Fig.~\ref{fig:pw-p}). Correspondingly, 
the static spin structure
factor calculated with respect to
${\cal P}_{S=0}
\big| \Psi_{\rm planar} \big\rangle$
\begin{equation}
S({\bm k}) = 
\sum_{\bm j} e^{i{\bm k}\cdot {\bm j}}C({\bm j})
\end{equation}
has characteristic
peaks at $(\pi,0)$ and $(0,\pi)$, while it
lose its weight at $(0,0)$ and $(\pi,\pi)$
(see Fig.~\ref{fig:FT}(a)).
This behavior resembles the structure factor
in the collinear antiferromagnetic phase and seems to
be consistent with a recent exact diagonalization
study up to $40$ sites.~\cite{rdsfr} In spite
of the prominent antiferromagnetic fluctuation
at $(0,\pi)$ and $(\pi,0)$, however,
the standard finite size scaling fitting of
the static spin structure factor~\cite{huse,richter}
suggests that the projected $Z_2$ planar
state does not have any finite sublattice
magnetization $M(\pi,0)$ in the
thermodynamic limit [see Fig.~\ref{fig:FT}(b)];
\begin{eqnarray}
[M(\pi,0)]^2=\lim_{L\rightarrow \infty} \frac{1}{L^2+2}
\sum_{{\bm j}}
C({\bm j})\!\ e^{i \pi j_x}  \simeq 0.
\end{eqnarray}

\begin{figure}
    \includegraphics[width=82mm]{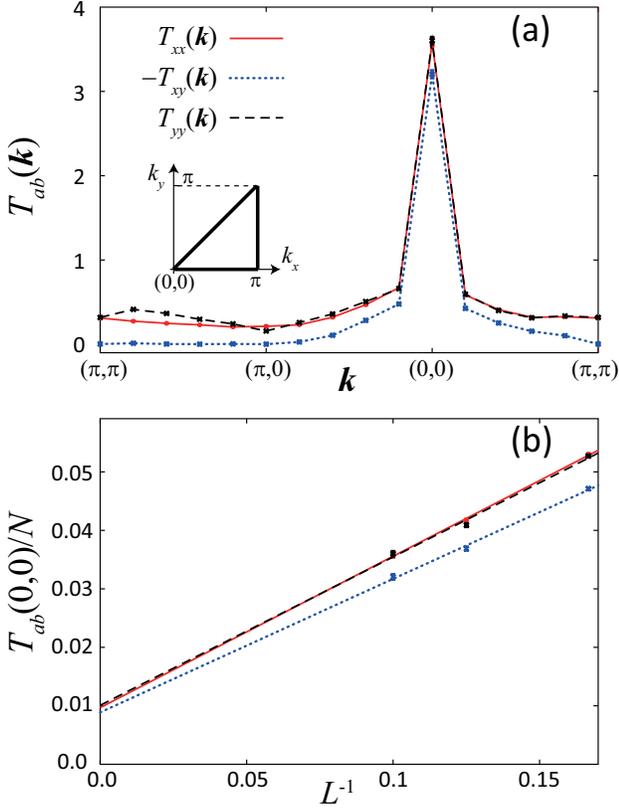}
\caption{(Color online) (a) Static structure factor
$ T_{ab}({\bm k})$ of the quadrupole moments
calculated in the projected $Z_2$ planar
state ${\cal P}_{S=0} |\Psi_{\rm planar}\rangle$,
with the same parameter set as used in
Fig.~\ref{fig:cr-p},
for the $J_1$--$J_2$ model with
$J_2=0.45 |J_1|$ in $10 \times 10$
spin cluster.  The (red) solid line depicts
the diagonal component $T_{xx}({\bm k})$,
the (blue) dotted line the off-diagonal
component with minus sign $-T_{xy}({\bm k})$,
and the (black) dashed line the other diagonal
component $T_{yy}({\bm k})$.
Inset:
The momentum ${\bm k}$
is taken along the high-symmetric
momentum points.
(b) Finite size scalings of
the long-range order of quadrupole moments $\frac{1}{N}
T_{ab}(0,0)$ as a function of $1/L$. The off-diagonal
component $T_{xy}(0,0)$ is shown multiplied by minus sign.
}
\label{fig:QQ-FT}
\end{figure}

The correlation functions of the quadrupole
moments are calculated in terms of
the projected $Z_2$ planar state in the
spin-singlet subspace,
\begin{align}
D_{ab}({\bm j}-{\bm m}) =& \sum^3_{\mu,\nu=1}
\langle \Psi_{\rm planar}|{\cal P}_{S=0} \!\
K^{\mu\nu}_{{\bm j},{\bm j}+{\bm e}_a} \nn \\
& \hspace{0.5cm} \times \!\
K^{\nu\mu}_{{\bm m},{\bm m}+{\bm e}_b} \!\
{\cal P}_{S=0} |\Psi_{\rm planar}\rangle \label{QQ2}
\end{align}
with $a,b=x,y$. We found that the diagonal
components thus calculated,
$D_{xx}({\bm j})$ and $D_{yy}({\bm j})$,
are positive-definite for any ${\bm j}$, while
the off-diagonal component always takes
a negative value, $D_{xy}({\bm j})<0$.
In the static structure factor of the
quadrupole moments
\begin{eqnarray}
T_{ab}({\bm k}) = 
\sum_{{\bm j}} e^{i{\bm k}\cdot{\bm j}}
D_{ab}({\bm j}), \label{FT-QQ}
\end{eqnarray}
both the diagonal components and
(minus of) the off-diagonal component exhibit
prominent peaks at the $\Gamma$-point
[see Fig.~\ref{fig:QQ-FT}(a)],
indicating the $d$-wave ordering
character of the quadrupole
moments in the projected $Z_2$ planar state.
In fact, finite size scalings of their peak
values suggest that the state is indeed
accompanied by finite quadrupole moments
in the thermodynamic limit
[see Fig.~\ref{fig:QQ-FT}(b)].

\section{summary and discussion}
In this paper, we have investigated the
phase diagram and the nature of a quantum spin
nematic phase
in the spin-$\frac12$ quantum frustrated
$J_1$--$J_2$ model with ferromagnetic $J_1$
on the square lattice,
describing the ground state wavefunction
in terms of a spin-triplet
pairing state of the spinon fields.
Our theory is based on the previous fermionic
mean-field analysis,\cite{sm}
which proposed four types of the mean-field solutions.
These solutions include (i)  $Z_2$ planar
state, (ii) $Z_2$ polar state, (iii) $SU(2)$
chiral $p$-wave state and
(iv) `flat-band' state, all of which are characterized
by different kinds of spin-triplet pairings of the
spinon fields introduced on ferromagnetic bonds.
Like in usual `projective description' of symmetric
quantum spin liquids,~\cite{wen,vmc,singlet,ma,Gros}
we construct projected BCS wavefunctions out
of these triplet pairing states.
Performing variational Monte Carlo simulations based
on these projected wavefunctions,
we obtain the phase diagram Fig.~\ref{fig:PD}
and static correlation
functions in the spin nematic wavefunction.

We first argue how these mean-field pairing
states are deformed, when external Zeeman
field is applied. A direct minimization of
the mean-field energy dictates that all the
$d$-vectors in these pairing states
are restricted within a plane perpendicular
to the applied field. This arrangement makes
these pairing states invariant under the spin
$\pi$-rotation around the magnetic field.
Owing to this $\pi$-rotational symmetry,
the corresponding projected
BCS wavefunctions acquire the `spin-nematic'
character; ordering of (the transverse component of)
the quadrupole moments without any spontaneous
ordering of (the transverse) spin moments nor
any magnetic crystallization. We
also show that the `flat-band' state,
which achieves the lowest mean-field energy
in the strong ferromagnetic regime $|J_1|\gg J_2$,
actually ends up in
the trivial fully polarized state, when projected
onto the spin Hilbert space.

To examine a possible realization of
quantum spin nematic states in the
present spin model, we study
the energetics of the projected (i)
$Z_2$ planar state, (ii)
$Z_2$ polar state, and (iii)
$SU(2)$ chiral $p$-wave state.
We focus especially on the intermediate
coupling regime, where
the ferromagnetic exchange $J_1$
is about twice as large as
the antiferromagnetic exchange $J_2$,
$J_2\simeq 0.5 |J_1|$. Based on the variational
Monte Carlo analysis, we argued that,
in a finite range of this intermediate
coupling regime, the projected
$Z_2$ planar state achieves the
best optimized energy, compared with
the energies of other competing phases,
such as the ferromagnetic phase
and collinear antiferromagnetic phase. (See Fig.~\ref{fig:PD}.)
We also prove that this projected
$Z_2$ planar state is accompanied by
the `$d$-wave' spatial configuration of the
quadrupole moments. This 
feature of the wavefunction including the irreducible
representations under the symmetry group
turns out to show a perfect agreement with the
nature of the quantum spin nematic
phase suggested by Shannon {\it et al.}\cite{sms} from
the exact diagonalization study.

Motivated by this coincidence,
we further calculate the
spin-spin correlation function of
the projected 
$Z_2$ planar state, so as
to obtain the static spin structure factor
in the spin nematic phase. 
The structure factor
thus calculated exhibits two prominent
peaks at the wavevectors
${\bm k}=(\pi,0)$ and $(0,\pi)$,
which signifies the presence of strong
collinear antiferromagnetic fluctuation.
The finite size scaling of the
peak height concludes that
the state does not possess any sublattice
magnetization in the thermodynamic limit,
which is consistent with the spin-nematic
feature of the $Z_2$ planar state.
This antiferromagnetic fluctuation is reminiscent
of the neighboring collinear antiferromagnetic phase,
whereas ferromagnetic fluctuation, which is also expected
to appear from the other neighboring ferromagnetic phase,
is completely suppressed.

These observations indicate that
the static spin-spin correlation function by itself
hardly distinguishes the current quantum spin
nematic phase from the neighboring
collinear antiferromagnetic phase.
Indeed, the recent exact
diagonalization study by
Richter {\it et al.}~\cite{rdsfr} reported
that the present $J_1$--$J_2$ model
exhibits only a strong collinear antiferromagnetic
correlation in the intermediate coupling regime,
$J_2 \simeq 0.5|J_1|$.
We expect that the dynamical magnetic
properties in combination with these static
physical properties could distinguish the
present spin nematic phase from the
collinear antiferromagnetic phase. That is, unlike in the
collinear antiferromagnetic phase, all the gapless Goldstone
modes in the spin-nematic phase are
expected to lose their spectral weight in the
dynamical spin structure factor in the
low-energy limit. In fact, a recent calculation
based on the random phase approximation
shows that this is indeed the case in
the $Z_2$ planar state.~\cite{sym}
This feature could be sharply contrasted to
the dynamical magnetic property in the collinear
antiferromagnetic phase, where the gapless mode at the
$(0,\pi)$-point or $(\pi,0)$-point
have a finite spectral weight even in
the low-energy limit.

\begin{acknowledgments}
We acknowledge Masatoshi Imada,
Yukitoshi Motome, Takashi Koretsune, Philippe Sindzingre,
and Leon Balents  for helpful discussions.
RS was partially supported by the Institute
of Physical and Chemical Research (RIKEN).
This
work was supported by Grants-in-Aid for Scientific Research
from MEXT, Japan (No.\ 22014016 and No.\ 23540397).
\end{acknowledgments}


\begin{thebibliography}{99}
\bibitem{comment}
In Ref.\ \onlinecite{sm}, we
called the planar state as the
(two-dimensional analogue of) Balian-Werthamer
state and the polar state as the collinear state.
In this paper, we adopt `planar/polar' instead,
since they seem to be
more familiar in the standard literature
such as Refs.\ \onlinecite{helium} and \onlinecite{helium2}.
\end{thebibliography}
\end{document}